\documentstyle[supertabular,epsf,rotate]{mn}

\newcommand{\mc}{\multicolumn}

\newcommand{\f}{\phantom{2}}
\newcommand{\ltsimeq}{\raisebox{-0.6ex}{$\,\stackrel 
        {\raisebox{-.2ex}{$\textstyle <$}}{\sim}\,$}} 
\newcommand{\gtsimeq}{\raisebox{-0.6ex}{$\,\stackrel 
        {\raisebox{-.2ex}{$\textstyle >$}}{\sim}\,$}} 

\input epsf.sty

\begin{document}

\title[The RLF from the low--frequency 3CRR, 6CE \& 7CRS samples] {The radio
luminosity function from the low--frequency 3CRR, 6CE \& 7CRS complete
samples}
\author[Willott et al.]{Chris J.\ Willott$^{1,2\star}$, Steve
Rawlings$^{1}$, Katherine M.\ Blundell$^{1}$, Mark Lacy$^{1,3,4}$ \and and
Stephen A. Eales$^{5}$\\ 
$^{1}$Astrophysics, Department of Physics, Keble Road, Oxford, OX1
3RH, U.K. \\
$^{2}$Instituto de Astrof\'\i sica de Canarias, C/ Via Lactea s/n,
38200 La Laguna, Tenerife, Spain \\
$^{3}$Institute of Geophysics and Planetary Physics, L-413 Lawrence
Livermore National Laboratory, Livermore, CA 94550, USA \\
$^{4}$Department of Physics, University of California, 1 Shields
Avenue, Davis CA 95616, USA\\
$^{5}$Department of Physics and Astronomy, University of Wales
Cardiff, P.O. Box 913, Cardiff CF2 3YB, U.K. \\ }

\maketitle

\begin{abstract}
\noindent
We measure the radio luminosity function (RLF) of steep-spectrum radio
sources using three redshift surveys of flux-limited samples selected
at low (151 \& 178 MHz) radio frequency, low--frequency source counts
and the local RLF. The redshift surveys used are the new 7C Redshift
Survey (7CRS) and the brighter 3CRR and 6CE surveys totalling 356
sources with virtually complete redshift $z$ information. This yields
unprecedented coverage of the radio luminosity versus $z$ plane for
steep-spectrum sources, and hence the most accurate measurements of
the steep-spectrum RLF yet made.  We find that a simple
dual-population model for the RLF fits the data well, requiring
differential density evolution (with $z$) for the two populations.
The low--luminosity population can be associated with radio galaxies
with weak emission lines, and includes sources with both FRI and FRII
radio structures; its comoving space density $\rho$ rises by about one
dex between $z \sim 0$ and $z \sim 1$ but cannot yet be meaningfully
constrained at higher redshifts.  The high--luminosity population can
be associated with radio galaxies and quasars with strong emission
lines, and consists almost exclusively of sources with FRII radio
structure; its $\rho$ rises by nearly three dex between $z \sim 0$ and
$z \sim 2$.  These results mirror the situation seen in X-ray and
optically-selected samples of AGN where: (i) low luminosity objects
exhibit a gradual rise in $\rho$ with $z$ which crudely matches the
rises seen in the rates of global star formation and galaxy mergers;
and (ii) the density of high luminosity objects rises much more
dramatically. The integrated radio luminosity density of the
combination of the two populations is controlled by the value of
$\rho$ at the low--luminosity end of the RLF of the high--luminosity
population, a quantity which has been directly measured at $z \sim 1$
by the 7CRS.  We argue that robust determination of this quantity at
higher redshifts requires a new redshift survey based on a large
($\sim 1000$ source) sample about five times fainter than the 7CRS.

\end{abstract}


\footnotetext{$^{\star}$Email: cjw@astro.ox.ac.uk}

\section{Introduction}
\label{sec:intro}

The radio luminosity function (RLF) seeks to derive from observed
samples and surveys of radio sources, their space density per unit
co-moving volume and how this changes with source luminosity. It is an
essential input to both gravitational lensing studies which probe
cosmic geometry (e.g. Kochanek 1996) and modelling of the clustering
of radio sources (Magliocchetti et al. 1998). The RLF derived at low
radio frequencies is important for understanding the content of high
frequency surveys and hence deriving jet beaming parameters
(e.g. Jackson \& Wall 1999). The shape and evolution of the RLF
provide important constraints on the nature of radio activity in
massive galaxies and its cosmic evolution.

Longair (1966) was one of the first to attempt to determine the
evolution of the radio source population. At that time there were very
few radio source redshifts known and the main constraint upon his
models came from the low--frequency source counts down to
$S_{151}\approx0.25$ Jy. Longair found that the data were best-fit by models
where the most powerful radio sources undergo greater cosmic evolution
(in comoving space density $\rho$) than less-powerful sources. The
subsequent acquisition of a substantial fraction of the redshifts for
the 3CRR complete sample of Laing, Riley \& Longair (1983) enabled the
evolution to be better constrained. Using the $V/V_{\rm max}$ test
(see Section \ref{vvmax}) they concluded that the most powerful radio
sources undergo substantial evolution with similar evolution for both
powerful radio galaxies and quasars. Unfortunately, the tight
correlation between radio luminosity and redshift in this sample meant
that they were unable to make any progress on the form of the evolving
RLF. Wall, Pearson \& Longair (1980) suggested that complete samples a
factor of $\approx 50$ times fainter than the 3CRR sample would be
necessary to differentiate between possible models of the RLF.

The most comprehensive study of the radio luminosity function prior to
this work was performed by Dunlop \& Peacock (1990; hereafter
DP90). Using several complete samples selected at 2.7 GHz with lower
flux-limits ranging from 2.0 Jy to 0.1 Jy, they considered the flat-
and steep-spectrum populations separately and derived the RLF for each
population. Their constraints on the RLF of flat-spectrum sources were
somewhat tighter than those derived by Peacock (1985), but the major
breakthrough of their paper concerned the constraints they were able
to place on the RLF of steep-spectrum radio sources. The strong
positive evolution in $\rho$ with $z$ out to $z \sim 2$ inferred by
Longair (1966), Wall et al. (1980) and Laing et al. (1983) was mapped
out in some detail, and more controversially their results indicated a
decline in the co-moving number density of both populations beyond a
redshift of $\approx 2$ -- the so-called `redshift cut--off'. DP90
concluded that for powerful sources a model of pure luminosity
evolution (PLE -- evolution only of the break luminosity with
redshift, analogous to that of Boyle, Shanks \& Peterson 1988 for the
quasar optical luminosity function) fitted the data well (but only for
$\Omega=1$). A similar model which incorporates negative density
evolution at high--redshift (the LDE model) was found to work well for
both $\Omega=1$ and $\Omega=0$. A recent update on the DP90 work can
be found in Dunlop (1998), and a critical re-evaluation of the
evidence for a redshift cut--off in the flat-spectrum population can
be found in Jarvis \& Rawlings (2000).

Derivations of the luminosity functions of luminous AGN selected at
all frequencies (radio, optical and X-ray) have shown a broken
power-law form with a steeper slope at the high--luminosity end than at
the low luminosity end (e.g. DP90, Boyle et al. 1988, Page et
al. 1996). These studies were also able to adequately describe the
positive evolution of the luminosity function from $z=0$ to $z \approx
2$ by assuming pure luminosity evolution. More recent studies with different
datasets have shown some deviations from PLE (e.g. Goldschmidt \&
Miller 1998, Miyaji, Hasinger \& Schmidt 2000). The physical rationale
behind PLE is that the AGN have long lifetimes (comparable with the
Hubble Time) and they decline in luminosity from $z \approx 2$ to
$z=0$. However, at least in the case of the high--luminosity radio
population, we know such a model is not viable because the ages of
FRII radio sources in the surveys are well-constrained to be
$\ltsimeq$ a few $10^8$ yr (e.g. Blundell, Rawlings \& Willott 1999).
Furthermore, for X-ray and optically-selected quasars, accretion at
the rates required for their high luminosities (over a Hubble Time)
would lead to much more massive black holes than are observed in the
local Universe (e.g. Cavaliere \& Padovani 1989). Thus, although PLE
models do fit the gross features of the data, the physical basis of
these models is not clear.

The radio source population is composed of several seemingly different
types of object. One division between types is concerned with radio
structure which falls into two distinct classes: FRI, or `twin-jet',
sources; and FRII, or `classical double', sources (Fanaroff \& Riley
1974). At low luminosities $ \log _{10} (L_{151}) \ltsimeq 25.5$, the
vast majority of sources have an FRI radio structure (apart from the
very low luminosity starbursts), whilst at higher luminosities the
majority of sources have FRII structure. A second division is
concerned with the optical properties of the population. At $\log
_{10} (L_{151}) \sim 26.5$ the fraction of objects in low--frequency
selected samples with observed broad-line nuclei changes rapidly from
$\approx 0.4$ at higher radio luminosities to $\approx 0.1$ at lower
luminosities (Willott et al. 2000).  It is not yet clear whether this
represents a fundamental change in central engine properties because,
for example, it could be due to the opening angle of an obscuring torus
increasing with decreasing luminosity (see Willott et al. 2000).
However, it is clear that at least some FRII sources lack even
indirect evidence for an active quasar nucleus because, as pointed out
by Hine \& Longair (1979) and Laing et al. (1994), they have weak or
absent emission lines. Indeed it is plausible that the majority of
FRII radio galaxies (per unit comoving volume) show passive
elliptical-galaxy spectra (Rixon, Wall \& Benn 1991) and are part of
the low--luminosity population considered here.

In this paper we have chosen to investigate the RLF in terms of a
dual-population model where the less radio-luminous population is
composed of FRIs and FRIIs with weak/absent emission lines, and the
more radio-luminous population of strong-line FRII radio galaxies and
quasars. Our motivation for doing this is that the presence/absence of
a quasar nucleus, as indicated by emission line strength, is a
distinction intimately connected to the properties of the central
engine, whereas radio structure is strongly influenced by the
larger-scale environment (e.g. Kaiser \& Alexander 1997).  The
critical radio luminosity [$\log _{10} (L_{151}) \sim 26.5$] which
crudely divides the populations is also very close to the break in the
RLF (e.g. from DP90), whereas the luminosity of the FRI/FRII divide is
clearly well below the break. Tentative evidence for an increase in
the fraction of quasars with redshift from $z=0$ to $z \approx 1$ can
then be explained if this low--luminosity population evolves less
rapidly than the high--luminosity population (Willott et al. 2000). The
$\rho$ of FRI sources is already known to evolve less-strongly with
$z$ than that of the radio-luminous FRIIs (e.g. Urry \& Padovani
1995), but here we will assume that this is because they are a sub-set
of a larger population of low--luminosity radio sources, including
FRIIs, which evolve less strongly.  We will show that differential
evolution between the two populations mimics PLE to a certain extent,
but with a more plausible physical basis.
 
Padovani \& Urry (1992), Urry \& Padovani (1995) and Jackson \& Wall
(1999) have adopted a different approach to determining the RLF from
high--frequency complete samples. Instead of separating the flat- and
steep-spectrum sources and deriving the RLF for each population, they
use models built around unified schemes and relativistic beaming to
determine the relative fractions of various classes of objects in the
samples. Their models assume that BL Lac objects are the beamed
versions of FRI and (in the case of Jackson \& Wall) weak-line FRII
radio galaxies, and that flat-spectrum quasars are favourably oriented
FRII radio galaxies.  However, both these analyses required the RLF of
unbeamed objects to be determined independently, i.e from
low--frequency samples. Because they used only the 3CRR sample to
determine the luminosity distribution of radio sources, the tight
$L_{178}-z$ correlation meant a degeneracy between luminosity and
redshift effects.  Current models of this type are thus most useful
for constraining the values of various beaming parameters and how
these affect the population mix in high--frequency selected samples,
and have not provided new information on the evolution of
steep-spectrum sources.

In this paper we present a new investigation of the steep-spectrum RLF
which combines the bright 3CRR sample with the new, fainter 6CE and
7CRS redshift surveys (based on the 6C and 7C radio surveys,
respectively). Note that an investigation of the RLF of steep-spectrum
quasars (based on the 3CRR and 7CRS samples) has been presented by
Willott et al. 1998 (W98). As the basis for an investigation of the
steep-spectrum RLF, the 3CRR/6CE/7CRS dataset improves on that used by
DP90 in four different ways.

\begin{itemize}

\item{ {\bf Spectroscopic completeness.}  Many of the sources in the
DP90 samples do not have spectroscopic redshifts [including, at the
time of DP90, $\approx50\%$ of the sources in their faintest sample --
the Parkes Selected Regions (PSR)], and redshifts had to be estimated
from the infrared Hubble diagram (the $K-z$ relation). Dunlop (1998)
reports that further spectroscopy of these sources has shown their
redshifts tended to be overestimated (due to the positive correlation
between radio luminosity and $K$-band luminosity, Eales et al. 1997),
which he contends strengthens the DP90 evidence for a redshift
cut--off in high--frequency selected samples. However, as we discuss
later (Section 5.1), the assignment of systematically high redshifts
to many sources may in fact have caused DP90 to overestimate the
strength of any redshift cut--off. In contrast to the PSR, the 6CE and
7CRS have virtually complete spectroscopic redshifts.}

\item{ {\bf Coverage of the radio luminosity versus $z$ plane.} The
flux-limit of the PSR is $S_{2.7}=0.1$ Jy. For steep-spectrum sources
with $\alpha = 0.8$, this corresponds to $S_{151} \approx 1$ Jy, and
for $\alpha = 1.2$ to $S_{151} \approx 3$ Jy, so the 7C Redshift
Survey (7CRS) with a flux-limit of $S_{151}=0.5$ Jy is a factor $2-6$
times deeper at a given $z$ than the PSR depending on spectral
index. Because of the correlations between $\alpha$ and radio
luminosity and $z$ (e.g. Blundell et al. 1999) this makes the 7CRS
sensitive to lower radio luminosity objects at the higher
redshifts. Note that because the PSR covers an area 3.5 times larger
than the 7CRS, the two samples actually contain similar numbers of
high--redshift objects.}

\item{ {\bf Radio completeness.} DP90 found the source counts in the
PSR at 0.1 Jy to be a factor of two lower than expected from the
source counts at other frequencies. They suggested that this is due to
incompleteness in the original survey near its flux-limit, although
this is by no means certain (J.V. Wall, priv. comm.). Although they
attempted to correct for this, there remains some possibility of bias:
methods based on the $V / V_{\rm max}$ statistic are particularly
sensitive to such problems, as objects close to the flux-limit have $V
/ V_{\rm max} \approx 1$, and losing such objects would bias the
statistic towards a spuriously low value.  The 6CE sample and the 7CRS
are based on surveys which are complete at the flux-densities of
interest. As Riley (1989) discusses, the number of large angular size
sources omitted from surveys such as these due to surface brightness effects
is likely to be negligible.}

\item{ {\bf Benefits of low--frequency selection}.  The ultimate goal
of our programme is to model the evolution in the RLF of
steep-spectrum sources. As detailed in Blundell et al. (1999), the
physics of FRII radio sources means that the mapping between radio
luminosity and one of the key physical variables, the bulk power in
the jets, is much closer if the radio luminosity is evaluated at low
rest-frame frequencies provided, of course, these lie above the
synchrotron self-absorption frequency of a given source. Modelling of
high frequency luminosity is much more problematic, both because lobe
emission at these frequencies is highly sensitive to source age and
redshift (because of inverse-Compton scattering of lobe electrons by
the microwave background), and because Doppler-boosted cores and
hotspots become much more important at high rest-frame frequencies. }

\end{itemize}

In Section 2 we discuss the data available to constrain the steep-spectrum
RLF. Section 3 describes the model fitting procedures and results. 
In Section 4 the evolution of the radio source
population is tested via the $V/V_{\rm max}$
statistic. Section 5 compares our findings with those of previous RLF
determinations. In Section 6 we summarise our results and indicate the
future work necessary to resolve the outstanding uncertainties.

The convention for the radio spectral index, $\alpha_{\rm rad}$, is
that $S_{\nu} \propto \nu^{-\alpha_{\rm rad}}$, where $S_{\nu}$ is the
flux-density at frequency $\nu$. We assume throughout that
$H_{\circ}=50~ {\rm km~s^{-1}Mpc^{-1}}$ and $\Omega_ {\Lambda}=0$. All
results are presented for both flat ($\Omega_ {\rm M}=1$) and open
($\Omega_ {\rm M}=0$) cosmologies.

\section{Constraining data}

The radio luminosity function (RLF) is defined as the number of radio
sources per unit co-moving volume per unit (base 10) logarithm of
luminosity, $\rho(L,z)$. To determine the RLF, we will make use of
three different types of radio data: complete samples with
spectroscopic redshifts; source counts; and a low--redshift sample to
fix the local RLF.

\begin{figure*}
\epsfxsize=0.9\textwidth
\epsfbox{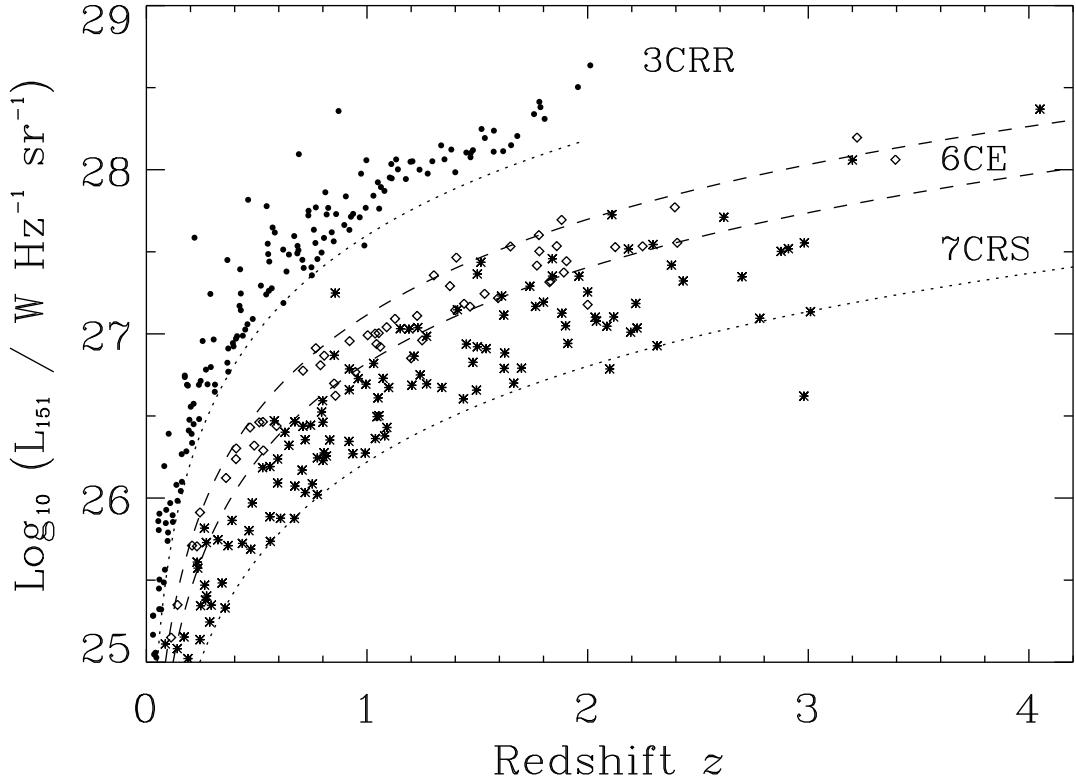}
{\caption[junk]{\label{fig:allpz} The radio luminosity--redshift plane
for the 3CRR, 6CE and 7CRS samples described in Section
\ref{compsamp}. The different symbols identify sources from different
samples: 3CRR (filled circles), 6CE (open diamonds) and 7CRS (asterisks).
The dotted lines show the flux-density lower limits for the 3CRR and 7CRS
samples and the dashed lines the limits for the 6CE sample (all assuming
a radio spectral index of 0.5). This plot is for $\Omega_ {M}=1$,
$\Omega_ {\Lambda}=0$.}}
\end{figure*}

\subsection{Complete samples}
\label{compsamp}

A complete sample consists of every radio source in a certain area on
the sky brighter than a specified flux-limit at the selection
frequency. If all these sources are identified optically with a galaxy
or quasar and redshifts are determined, then the distribution of
points on the $L-z$ plane, together with knowledge of the sample
flux-limits and sky areas, in principle allow the determination of the
RLF. The major uncertainty in such a determination is that at least
some regions of the $L-z$ plane will be very sparsely populated, and
at these $L$ and $z$ the RLF measurement will be prone to small number
statistical uncertainties. These typically arise because the
identification of and/or redshift acquisition of faint radio sources
can be very time-consuming, and due to the limited nature of time
available at large telescopes, complete samples up until now have
generally been limited to $\ltsimeq 100$ sources.  Some regions of the
$L-z$ plane are prone to a more fundamental problem: objects of a
given $L$ may be sufficiently rare that there is simply too little
observable comoving volume at the redshift of interest to obtain large
samples.

Our dataset of redshift surveys combines three low--frequency complete
samples with different flux-limits, which together provide
unparalleled coverage of the $L_{151}-z$ plane (see
Fig. \ref{fig:allpz}).

\subsubsection{7CRS} 

The faintest of the three complete samples used is the new 7C Redshift
Survey (7CRS). This sample contains every source with a low--frequency
flux-density $S_{151} \geq 0.5$ Jy in three regions of sky covering a
total of 0.022 sr. The regions 7C-I and 7C-II will be described in
Blundell et al. (in prep.) and 7C-III in Lacy et al. (1999a) and
references therein.  For 7C-I and 7C-II, multi-frequency radio data
including high--resolution radio maps from the VLA will be presented in
Blundell et al. From measurements at several frequencies, the radio
spectra have been fitted and hence radio luminosities at rest-frame
151 MHz determined. Spectral indices at various frequencies were also
determined from these fits. For 7C-III, the availability of 38 MHz
data from the 8C survey in the same region has allowed the
low--frequency spectral indices to be calculated and hence rest-frame
151 MHz luminosities (Lacy et al. 1999a).

Due to the low--frequency selection of the sample, it is expected to
contain few flat-spectrum ($\alpha_{1{\rm GHz}} <0.5$) sources. From
inspection of the VLA maps and radio spectra, sources which only have
flux-densities above the sample limit due to their Doppler-boosted
cores can be identified. Out of the total of 130 sources, only one
source fulfills these criteria -- the quasar 5C7.230. Therefore this
source is excluded from the analysis presented in this paper. Note
however that there are five other quasars which have $\alpha_{1 {\rm
GHz}}<0.5$, the traditional separation value between flat- and
steep-spectrum quasars. All of these have $\alpha_{1 {\rm GHz}}\approx
0.4$ and are compact ($\theta<5$ arcsec), but have extended emission
on {\em both} sides of the core, indicating that Doppler-boosting is
not the cause of this emission. They all have sufficient extended flux
to remain in the sample and as such they are not excluded on
orientation bias grounds.  These sources are referred to as Core-JetS
sources (CJSs; see Blundell et al. in prep. for more details). The
one other exclusion is 3C200, because it is already included in the
3CRR sample, leaving a total of 128 sources.

All 7C Redshift Survey sources have been reliably identified with an
optical/near-infrared counterpart. For 7C-I and 7C-II, spectroscopy
has been attempted on all sources and secure redshifts obtained for $>
90$\% of the sample (Willott et al., in prep.). Seven sources do not
show any emission or absorption lines in their spectra. For these
objects, we have obtained multi-colour optical and near-infrared
photometry (in the $R$, $I$, $J$, $H$ and $K$ bands) to attempt to
constrain their redshifts. All of these objects have spectral energy
distributions consistent with evolved stellar populations at redshifts
in the range $1 \leq z \leq 2$ (Willott, Rawlings \& Blundell 2000). These
derived redshifts are generally consistent with those from the $K-z$
diagram, particularly given the increased scatter in the relationship
at faint magnitudes (Eales et al. 1997). Note that radio galaxies in
the redshift range $1.3<z<1.8$ are traditionally difficult to obtain
redshifts for because there are no bright emission lines in the
optical wavelength region. The correlation between emission line and
radio luminosities (Baum \& Heckman 1989; Rawlings \& Saunders 1991;
Willott et al. 1999) exacerbates this problem for faint radio
samples. We believe that this is the primary reason why we have been
unable to detect emission lines in these 7 sources. It is extremely
unlikely that any of these objects actually lie at $z>2.5$.

Details of the imaging and spectroscopy in the 7C-III region are given
in Lacy et al. (1999a, 1999b, 2000). Reasonably secure spectroscopic
redshifts have been obtained for 81\% of the sources and uncertain
redshifts for a further 13\%. Near-infrared magnitudes recently
obtained support these uncertain redshifts. Only three sources have no
redshift information. Near-infrared and optical imaging shows that two
of them have SEDs typical of $z \approx 1.5$ galaxies as is the case
for those without spectroscopic redshifts in 7C-I and 7C-II. The other
object (7C1748+6703) has a break between $J$ and $K$ that is
suggestive of a redshift $z>2.4$. The existence of significant flux in
its optical spectrum down to 6000 \AA~ and a $K$-magnitude of 18.9,
suggest $z<4$. Therefore we adopt $z=3.2$ in this paper.

\subsubsection{3CRR}

The bright complete sample used is the 3CRR sample of Laing, Riley \&
Longair (1983, hereafter LRL) with complete redshift information for
all 173 radio sources. The flux-limit is 10.9 Jy at 178 MHz, which
translates to 12.4 Jy at 151 MHz assuming a typical spectral index of
0.8. Blundell et al. (in prep.) review the current status of this
sample and presents the results of spectral fitting of multi-frequency
radio data. 3C231 (M82) is excluded because the radio emission from
this source is due to a starburst and not an AGN. The two
flat-spectrum quasars, 3C345 and 3C454.3, are excluded because only
Doppler-boosting of their core fluxes raises their total fluxes above
the flux-limit. To include objects such as these would cause an
overestimate in the number density of AGN, since they only get into
the samples because their jet axes are very close to our
line-of-sight. Note that high--frequency selected samples contain a
much higher proportion of objects such as these, showing that our
choice of low--frequency selection virtually eliminates these
favourably oriented sources [there are only three such sources
($\approx 1$\%) in our complete samples].

\subsubsection{6CE}

The 6CE complete sample we use is a revision of the 6C sample of Eales
(1985) (see Rawlings, Eales \& Lacy 2000 for details). The flux-limits
of this sample are $2.0\leq S_{151}<3.93$ Jy and the sky area covered
is 0.103 sr. Only 3 of the 59 sources in this sample do not have
redshifts determined from spectroscopy. One object is occluded by a
bright star and is excluded from further consideration (without
bias). One object is faint in the near--infrared ($K > 19$), so we
take it as a galaxy at $z=2.0$ in this paper. The other is relatively
bright in the near-IR but very faint in the optical. Its red colour
suggests a galaxy with a redshift in the range $0.8<z<2$ and we assume
$z=1.4$ here. None of the 6CE sources are clearly promoted into the
sample by Doppler-boosted core emission (c.f 5C7.230, 3C345 and
3C454.3), so all 58 sources are included in our analysis. Full details
of the sample, including optical spectra, are given in Rawlings et
al. (2000).

\subsection{The source counts at 151 MHz}

The three complete samples with known redshift distributions contain a
total of 356 sources (after the exclusions mentioned above). However,
the number of radio sources at 151 MHz as a function of flux density,
also known as the source counts, has been determined for thousands of
sources from the much larger sky area 6C and 7C surveys (Hales,
Baldwin \& Warner 1988 and McGilchrist et al. 1990, respectively). The
7C source counts go as faint as 0.1 Jy -- a factor of five fainter
than the 7CRS.  At these flux-densities the source counts are our only
constraint on the RLF. At the bright end ($\gtsimeq 10$ Jy), the
source counts are obtained by binning the sources in the 3CRR sample
of LRL in 5 flux bins. The three sources excluded in Section
\ref{compsamp} were also excluded for the calculation of the source
counts. The 3CRR source counts are converted from 178 to 151 MHz by
assuming a spectral index of 0.8.

\begin{figure}
\epsfxsize=0.47\textwidth
\epsfbox{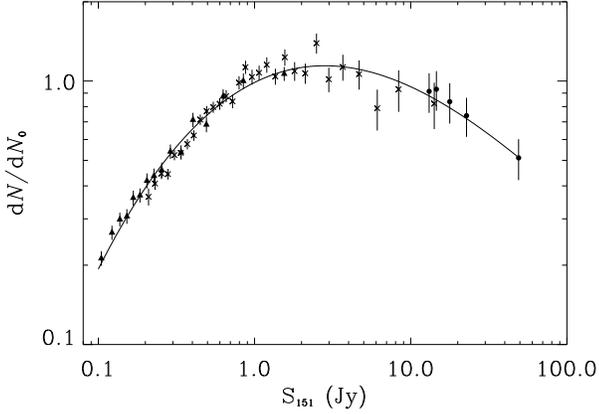}
{\caption[junk]{\label{fig:polysc} The binned differential source
counts from the 3CRR (LRL), 6C (Hales et al. 1988) and 7C (McGilchrist
et al. 1990) surveys.  Error bars are $\sqrt N$ Poisson errors for the
number of sources in each bin. The symbols identify the surveys: 3CRR
(circles), 6C (crosses) and 7C (triangles). The line shows the third-order 
polynomial fit to the data points.}}
\end{figure}

Figure \ref{fig:polysc} shows the binned differential source counts
from the 3CRR, 6C and 7C surveys. The counts are normalised to the
differential source counts for a uniform distribution in a Euclidean
universe, such that $dN_{0}=2400 \left( S_{\rm min}^{-1.5}
-S_{\rm max}^{-1.5} \right)$, where $S_{\rm min}$ and
$S_{\rm max}$ are the lower and upper flux limits of the
bin. These points were fitted by a third-order polynomial using a
least-squares fit in logarithm space, i.e. 
\begin{equation}
\log_{10} \left( \frac{dN}{dN_{0}} \right) = \begin{array}{l} a_{0} + a_{1}\log_{10} S_{151} + \\ a_{2} \left( \log_{10} S_{151} \right) ^{2} + a_{3} \left( \log_{10} S_{151} \right) ^{3}.
\end{array}
\end{equation}
The coefficients of the fit are $a_{0}=-0.00541, a_{1}=0.293,
a_{2}=-0.362$ and $a_{3}=0.0527$. A higher order polynomial was not
required to provide a decent fit to the data. The error on this curve
was estimated using the errors from the source count bins. At $S_{151}
\leq 1.0$ Jy the fractional errors on the bins are approximately
constant with flux-density at the level of 0.06. At $S_{151} > 1.0$ Jy the
fractional errors are given by a power-law with slope 0.49 such that
the fractional error at 10 Jy is 0.19, for example.

It is well known that there is a flattening of the source counts slope
at 1.4-GHz flux-densities $\sim 1$ mJy. This is attributed to the
counts becoming dominated by low--redshift star-forming galaxies
(Condon 1984; Windhorst et al. 1985).  Although there is only one such
starburst galaxy in our complete samples down to 0.5 Jy (3C231 ), the
source count data goes as low as 0.1 Jy, so it is essential to
investigate the contribution of starburst galaxies to the source
counts at these low flux-densities. To estimate the starburst
contribution we used the $60 \mu$m luminosity function of Saunders et
al. (1990) scaled by the mean of the ratio of IRAS $60 \mu$m to NVSS
1.4 GHz flux-densities (Cotton \& Condon 1998). Cosmic evolution is
accounted for by the PLE model adopted by Saunders et al.  To
translate this 1.4 GHz luminosity function to 151 MHz, a spectral
index of 0.7 was assumed for the starburst population. The
differential source counts at 151 MHz were then calculated from this
luminosity function. It is found that the starburst contribution at
0.1 Jy is only 2\% of the total source counts. This is within the
uncertainty in the source counts at 0.1 Jy, so it can be safely
neglected.

\subsection{The local RLF} 

The radio luminosity function at low--redshift ($z\ltsimeq 0.2$) is
fairly well-determined. This therefore provides a boundary condition
for determination of the evolving RLF. Unfortunately, the low space
density of powerful radio sources and small cosmological volume out to
$z\approx 0.2$ means that the local RLF (LRLF) is only
well-constrained at radio luminosities below $L_{151} \approx 10^{24}$
W Hz$^{-1}$ sr$^{-1}$. Hence the LRLF will only constrain the faint
end of the RLF. Another problem is that at very low luminosities
($L_{151} \approx 10^{21}$ W Hz$^{-1}$sr$^{-1}$) the LRLF is dominated
by the star-forming spiral/irregular galaxies which dominate the
sub-mJy (at 1.4 GHz) 
source counts. Since we do not want to include this population
in our RLF determination, it is essential that the LRLF of the AGN
population only is used.

A determination of the local radio luminosity function which treats
AGN and star-forming galaxies as two distinct populations is given by
Cotton \& Condon (1998). By cross-matching UGC galaxies (Nilson 1973)
with the NVSS catalogue (Condon et al. 1998) they derived luminosity
functions at 1.4 GHz for each population. The AGN LRLF at 151 MHz was
derived from this by assuming a radio spectral index of 0.8. This
determination of the LRLF by Cotton \& Condon is well-approximated by
a power-law (up to $L_{151} \approx 10^{24}$W Hz$^{-1}$ sr$^{-1}$)
with an index of $-0.53$.  From this model the LRLF was determined in
10 bins in the range $20\leq {\rm \log}_{10} (L_{151} /$ W Hz$^{-1}$
sr$^{-1}) \leq 23.6$. The error on each bin was assumed to be 0.1 dex.

\section{Model fitting} 
\label{rlfmod}
\subsection{Method} 

To find best-fit parameters for various models of the evolving RLF, a
maximum likelihood method is adopted. This method is similar to that
used in W98 to determine the quasar RLF and we refer readers to that
paper for further details. The aim of the maximum likelihood method is
to minimise the value of $S$, which is defined as
\begin{displaymath}
S = -2 \sum^{N}_{i=1} \ln [\rho(L_{i},z_{i})]+
\end{displaymath}
\begin{equation}
\label{eqn:qlike}
\f \f \f \f \f 2 \int \hspace{-0.25cm} \int \rho(L,z) 
\Omega(L,z) \frac{dV}{dz}  dzd \log_{10} L,
\end{equation}
where $(dV/dz)dz$ is the differential co-moving volume element,
$\Omega(L,z)$ is the sky area available from the samples for these
values of $L$ and $z$ and $\rho(L,z)$ is the model distribution being
tested. In the first term of this equation, the sum is over all the
$N$ sources in the combined sample. The second term is simply the
integrand of the model being tested and should give $\approx 2N$ for
good fits.

The only difference here is that the source counts and local radio
luminosity function provide additional constraints for the fitting
process. The source counts and LRLF are one-dimensional functions and
to estimate how well they fit the model, the value of $\chi^{2}$ is
evaluated thus:
\begin{equation}
\chi^{2}=\sum^{N}_{i=1} \left( \frac {f_{{\rm data}~ i} - f_{{\rm
mod}~ i}} {\sigma_{{\rm data}~ i}} \right) ^{2},
\end{equation}
where $f_{{\rm data}~ i}$ is the value of the data in the $i$th bin,
similarly $f_{{\rm mod}~ i}$ and $\sigma_{{\rm data}~ i}$ are the
model value and data error in the $i$th bin, respectively. These are
all determined in logarithm space for both the source counts and the
LRLF. The sum is over the $N$ bins in each case.

In order to combine the constraints from all three types of data, we
follow Kochanek (1996). $\chi^{2}$ is related to the likelihood by
$\chi^{2}=-2 \ln$ (likelihood), i.e. the same form as $S$. Therefore
the function that is now minimised is
\begin{equation}
S_{\rm all}= \chi_{\rm SC}^{2} + \chi_{\rm LRLF}^{2} + S - S_{0},
\end{equation}
where $S_{0}$ is a constant which normalises $S$ so that equal
statistical weight is given to all three types of data. The value of
$S_{0}$ theoretically comes from the dropped terms from equation
\ref{eqn:qlike}. However, these are not possible to calculate
directly, so instead one estimates the value of $S_{0}$ so that the
first term in equation \ref{eqn:qlike} equals $\approx 0$ and $S
\approx 2N$. Errors on the best-fit parameters are 68\% confidence
levels, as discussed in Boyle et al. (1988).

The goodness-of-fit of models are estimated using the 2-D
Kolmogorov-Smirnov (KS) test (Peacock 1983), as in W98. Note that it
is expected that the KS probabilities will be lower than for the
quasar RLF fits, because in general the probability decreases with
increased sample size in this test. Simulations of the $L_{151}-z$
plane are used to double check the goodness-of-fit. In addition, the
predicted redshift distributions for the complete samples are derived
from the models and compared with the actual distribution. This is
particularly useful in testing for the presence of a redshift cut--off
at high--$z$.

\subsection{The form of the RLF} 
\label{rlfform}

As explained in Sec.~\ref{sec:intro}, the RLF is modelled here as a
combination of two populations which are allowed to have different RLF
shapes and evolutionary properties. The low--luminosity radio source
population (composed of FRIs and low--excitation/weak emission line
FRIIs) is modelled with a Schechter function,
\begin{equation}
\rho_{\rm l}(L) =\rho_{{\rm l}\circ} \left( \frac{L}{L_{{\rm l} \star}} \right) ^{-\alpha_{\rm l}} \exp  {\left( \frac{-L}{L_{{\rm l} \star}} \right) }, 
\end{equation}
where $\rho_{\rm l}(L)$ is the source number density as a function of
radio luminosity $L$, $\rho_{{\rm l}\circ}$ is a normalisation term,
$L_{{\rm l} \star}$ is the break luminosity and $\alpha_{\rm l}$ is
the power-law slope. The evolution function of $\rho_{\rm l}$ is
modelled as $f_{\rm l} (z) = (1+z)^{k_{\rm l}}$ up to a maximum
redshift $z_{{\rm l}\circ}$ beyond which there is no further
evolution. Hence there are five free parameters to be fixed for the
low--luminosity population.

For the high--luminosity population a similar form is adopted, except
in this case the exponential part of the Schechter function is at low
luminosity and the power-law region is at high luminosity, i.e.
\begin{equation}
\rho_{\rm h}(L) =\rho_{{\rm h}\circ} \left( \frac{L}{L_{{\rm h} \star}} \right) ^{-\alpha_{\rm h}} \exp {\left( \frac{-L_{{\rm h} \star}}{L} \right) }, 
\end{equation}
where the subscript h refers to the high luminosity population. Note
that it is expected that $L_{{\rm l} \star} \approx L_{{\rm h}
\star}$, so that at this luminosity the decline in one population is
compensated by the rise of the other, since no discontinuity or
bi-modality in the RLF has yet been observed (e.g. DP90). These forms
for the luminosity functions of the two populations were chosen
because the steep exponentials ensured a fairly small overlap in
$L_{151}$ between the two populations. However, we currently have
little information about the relative population fractions in the
overlap region, so this is just a convenient method of separating the
populations. We note that the low-luminosity RLF of our model has a
similar shape to the elliptical galaxy $K$-band luminosity (and
therefore mass) function (e.g. Gardner et al. 1997).

The high--luminosity population is assumed to undergo density
evolution of a similar form to that of the quasar RLF in W98. The
$z$-distribution $f_{\rm h} (z)$ is given by a Gaussian in redshift
(model A here), with a variant on this distribution being a one-tailed
Gaussian from zero redshift to the peak redshift and then a constant
density to high--redshift (model B -- also known as the `no cut--off'
model since there is no redshift cut--off). Note that the Gaussian
model A is a convenient approximation to the shape of the evolution of
luminous optically-selected quasars (e.g. Warren, Hewett \& Osmer
1994), and to the evolution inferred for flat-spectrum radio sources
by Shaver et al. (1999) -- but see Jarvis \& Rawlings (2000) for a
contrary view.  There is no physical reason why one would expect the
evolution to be symmetric in redshift. In W98 we showed that the
Gaussian redshift distribution from $z=0$ to the peak is very similar
to the form $(1+z)^{k}$, so we do not include models with this form of
evolution here.

In W98 it was found that both these redshift distributions (models A and B in this paper) provided
good fits because the number of quasars in the samples at
high--redshift was fairly small. The inclusion of the 6CE and 7C-III
samples and the 7CRS radio galaxies gives a much larger sample of
$z>2$ objects here than for the quasar RLF. Hence, the high redshift
evolution can be better determined. Therefore, an additional redshift
distribution is tested here, which has a one-tailed Gaussian rise to
the peak redshift and then a one-tailed Gaussian decline at higher
redshifts which is allowed to have a different width from the
rise. This allows more shallow high--redshift declines in $\rho_{\rm
h}$ (such as that inferred by DP90) than the symmetric model A.
This model (C) directly fits the strength of any decline in the
co-moving density beyond the peak redshift $z_{{\rm h}\circ}$, since
the value of the high--redshift Gaussian width $z_{{\rm h}2}$ is free
to be fit from 0 for an abrupt cut--off to a large value for a very
gradual decline at high--redshift. Hence there are a total of six free
parameters in the high--luminosity population model with this form of
redshift evolution. 

To summarise our adopted RLF models, below we give all the equations
needed to reproduce the RLF $\rho(L,z)$. The values of the best-fit
parameters are given in Table \ref{tab:rlffit}.
\begin{equation}
\rho(L,z) =\rho_{\rm l} +\rho_{\rm h} 
\end{equation}
where
\begin{equation}
\rho_{\rm l} =\rho_{{\rm l}\circ} \left( \frac{L}{L_{{\rm l} \star}} \right) ^{-\alpha_{\rm l}} \exp  {\left( \frac{-L}{L_{{\rm l} \star}} \right) } (1+z)^{k_{\rm l}}  ~~~~~ {\rm for}~ z<z_{{\rm l}\circ},
\end{equation}
\begin{equation}
\rho_{\rm l} =\rho_{{\rm l}\circ} \left( \frac{L}{L_{{\rm l} \star}} \right) ^{-\alpha_{\rm l}} \exp  {\left( \frac{-L}{L_{{\rm l} \star}} \right) } (1+z_{{\rm l}\circ})^{k_{\rm l}} ~~~ {\rm for}~ z\geq z_{{\rm l}\circ},
\end{equation}
\begin{equation}
\rho_{\rm h} =\rho_{{\rm h}\circ} \left( \frac{L}{L_{{\rm h} \star}} \right) ^{-\alpha_{\rm h}} \exp  {\left( \frac{-L_{{\rm h} \star}}{L} \right) } f_{\rm h} (z). 
\end{equation}
The high--luminosity evolution function  $f_{\rm h} (z)$ has three different forms depending upon the model and the redshift:
\begin{equation}
f_{\rm h}(z) = \exp \left\{ - \frac {1}{2} {\left( \frac{z-z_{{\rm h}\circ}}{z_{{\rm h}1}} \right)^{2} } \right\}
\end{equation}
for model A at all redshifts and models B and C at $z<z_{{\rm h}\circ}$;
\begin{equation}
f_{\rm h}(z) = 1.0
\end{equation}
for model B at $z\geq z_{{\rm h}\circ}$;
\begin{equation}
f_{\rm h}(z) = \exp \left\{ - \frac {1}{2} {\left( \frac{z-z_{{\rm h}\circ}}{z_{{\rm h}2}} \right)^{2} } \right\}
\end{equation}
for model C at $z\geq z_{{\rm h}\circ}$. 

\begin{table*}
\footnotesize
\begin{center}
\begin{tabular}{ccccccccccccc}
\hline\hline 
\mc{1}{c}{Model} &\mc{1}{c}{$\Omega_{\rm M}$} &\mc{1}{c}{$\log(\rho_ {{\rm l}\circ})$}&\mc{1}{c}{$\alpha_{\rm l}$}&\mc{1}{c}{$\log(L_{{\rm l} \star})$}&\mc{1}{c}{$z_{{\rm l}\circ}$}&\mc{1}{c}{$k_{\rm l}$}&\mc{1}{c}{$\log$ ($\rho_ {{\rm h}\circ}$)}&\mc{1}{c}{$\alpha_{\rm h}$}&\mc{1}{c}{$\log(L_{{\rm h} \star})$}&\mc{1}{c}{$z_{{\rm h}\circ}$}&\mc{1}{c}{$z_{{\rm h}1}$}&\mc{1}{c}{$z_{{\rm h}2}$}\\

\hline\hline

A & 1 & $-7.153$ & $0.542$ & $26.12$ & $0.720$ & $4.56$ 
  & $-6.169$ & $2.30$  & $27.01$ & $2.25$  & $0.673$ & -- \\

B & 1 & $-7.150$ & $0.542$ & $26.14$ & $0.646$ & $4.10$
  & $-6.260$ & $2.31$  & $26.98$ & $1.81$  & $0.523$ & -- \\

C & 1 & $-7.120$ & $0.539$ & $26.10$ & $0.706$ & $4.30$ 
  & $-6.196$ & $2.27$  & $26.95$ & $1.91$  & $0.559$ & $1.378$ \\
\hline
A & 0 & $-7.503$ & $0.584$ & $26.46$ & $0.710$ & $3.60$ 
  & $-6.740$ & $2.42$  & $27.42$ & $2.23$  & $0.642$ & -- \\

B & 0 & $-7.484$ & $0.581$ & $26.47$ & $0.580$ & $3.11$ 
  & $-6.816$ & $2.40$  & $27.36$ & $1.77$  & $0.483$ & -- \\

C & 0 & $-7.523$ & $0.586$ & $26.48$ & $0.710$ & $3.48$
  & $-6.757$ & $2.42$  & $27.39$ & $2.03$  & $0.568$ & $0.956$ \\

\hline\hline         
\end{tabular}\\
 Errors for model C ($\Omega_{\rm M}=1$):~ $\log(\rho_ {{\rm l}\circ})$; $-0.11, +0.10$,~ $\alpha_{\rm l}$; $-0.02, +0.02$,~ $\log(L_{{\rm l} \star})$; $-0.09, +0.08$,~ $z_{{\rm l}\circ}$; $-0.10, +0.10$,~ $k_{\rm l}$; $-0.55, +0.57$,~ $\log$ ($\rho_ {{\rm h}\circ}$); $-0.11, +0.09$,~ $\alpha_{\rm h}$; $-0.11, +0.12$,~ $\log(L_{{\rm h} \star})$; $-0.10, +0.11$,~ $z_{{\rm h}\circ}$; $-0.16, +0.16$,~ $z_{{\rm h}1}$; $-0.05, +0.05$,~ $z_{{\rm h}2}$; $-0.28, +0.52$.

\end{center}              
{\caption[Table of observations]{\label{tab:rlffit}Best-fit parameters
for RLF models A, B and C, as described in Section \ref{rlfform}. The
top three rows refer to $\Omega_{\rm M}=1$ and the bottom three rows
$\Omega_{\rm M}=0$. }}
\normalsize
\end{table*}

In this paper, we have performed the RLF modelling for two
cosmological models with $\Omega_ {\rm M}=1$ and $\Omega_ {\rm M}=0$
and zero cosmological constant. To obtain a fairly reliable estimate
of the RLF in other cosmologies one can use the following relation
from Peacock (1985),
\begin{equation}
\label{eqn:peacosmo}
\rho_1(L_{1},z)\frac{dV_1}{dz}=\rho_2(L_{2},z)\frac{dV_2}{dz},
\end{equation}
where $L_{1}$ and $L_{2}$ are the luminosities derived from the
flux-density and redshift in the two cosmologies. We have tested this
relation by converting our model C RLF derived for $\Omega_ {\rm M}=0$
back to $\Omega_ {\rm M}=1$ and then compared this with the RLF
derived for $\Omega_ {\rm M}=1$. We note that the relation works well
for regions of the luminosity function that are well--constrained by
data. However, in regions which are constrained by little or no data,
the relation may not reproduce a model RLF close to that which would
emerge from direct fitting to the data in the new cosmology. Thus,
Table \ref{tab:rlffit} shows that the model C peak redshifts $z_{{\rm
h}\circ}$ and strength of high-$z$ decline $z_{{\rm h}2}$ are
different for $\Omega_ {\rm M}=1$ and $\Omega_ {\rm M}=0$ and equation
\ref{eqn:peacosmo} cannot reproduce such differences. Note that
the RLF for $\Omega_ {\rm M}=0.3$, $\Omega_ {\Lambda}=0.7$ is very
similar to that of $\Omega_ {\rm M}=0$, $\Omega_ {\Lambda}=0$ for
regions of the RLF well--constrained by our data and equation
\ref{eqn:peacosmo} can be applied in this case.

\subsection{Results} 
\label{results}

\begin{figure*}
\epsfxsize=0.95\textwidth \epsfbox{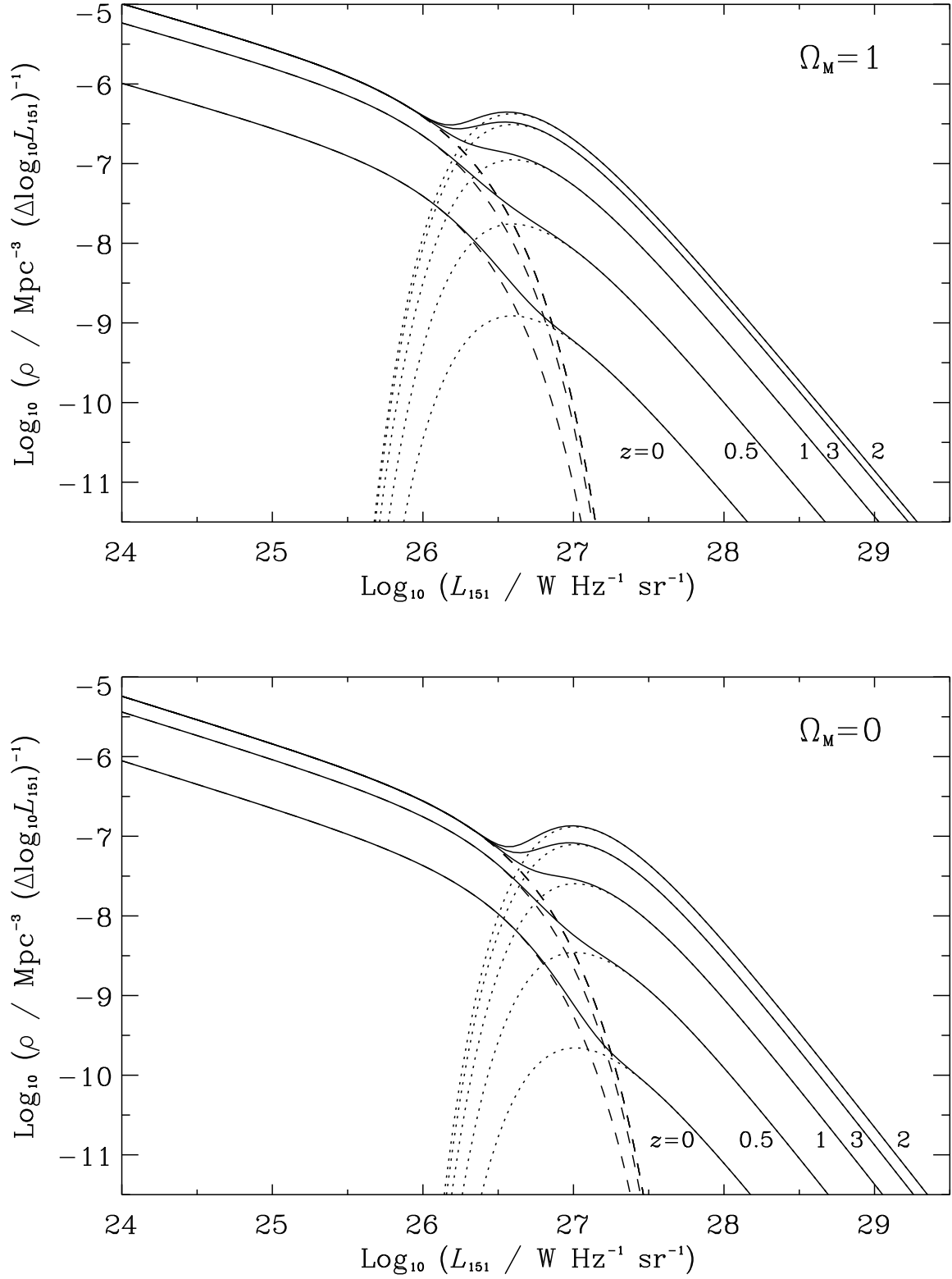}
{\caption[junk]{\label{fig:allrlf} The radio luminosity function
derived for model C for $\Omega_{\rm M}=1$ (top panel) and $\Omega_{\rm
M}=0$ (bottom panel). Dashed lines show the low--luminosity population
at $z=0, z=0.5$ and $z\gtsimeq 0.7$ (bottom to top). The dotted lines
are the high--luminosity RLF at $z=0,0.5,1,3$ and $2$ (bottom to
top). Solid lines show the sum of both components.}}
\end{figure*}

The maximum likelihood fitting routine was run for these three RLF models.
The routine integrated over the ranges $ 20 < \log_{10} (L_{151}) <
30$ and $0<z<5$. A check was made to ensure that the same results
occurred if the redshift range was increased to $z=10$. All three models
provided good fits to the data. The best-fit parameters for each model
are shown in Table \ref{tab:rlffit}. 

The parameters for the low--luminosity population RLF do not change
significantly for the different models, which is as expected, since it
has little to do with the form of the high--$z$ evolution of the
high--luminosity RLF. Note that the value of the break luminosity,
$\log(L_{{\rm l} \star}) \approx 26.1$ (26.5 for $\Omega_{\rm M}=0$),
is above the FRI/FRII divide and below the break in the quasar RLF
from W98 [$\log(L_{\rm break}) \approx 26.8$ (27.2 for
$\Omega_{\rm M}=0$)]. The models show significant evolution ($k_{\rm l}
\approx 4$) for the low--luminosity objects out to $z\approx0.7$ with
no further evolution to higher redshifts. Although it is often stated
that the FRI radio galaxy population shows no evolution, the samples
used to determine this generally have small numbers of objects and a
narrow redshift range (e.g. Urry, Padovani \& Stickel 1991; Jackson \&
Wall 1999). As we shall see in Section \ref{vvmax} the combination of
the 3CRR, 6CE and 7CRS samples shows considerable low--redshift evolution
for low--luminosity ($\log_{10} L_{151} \ltsimeq 26.5$) FRI and FRII
radio galaxies.

The best-fit models for the high--luminosity RLF have much steeper
slopes than the low--luminosity RLF ($\alpha_{\rm h}\approx2.3$, c.f.
$\alpha_{\rm l}\approx0.6$). This, together with $\log(L_{{\rm l}
\star}) \sim \log(L_{{\rm h} \star})$, gives the whole RLF a broken
power-law form, such as observed by DP90 for the RLF and by Boyle et
al. (1988) for the quasar OLF. $L_{{\rm h} \star}$ in these models is
about an order of magnitude greater than $L_{{\rm l} \star}$ and is
very similar to the break in the quasar RLF estimated in W98 from
source counts constraints. The RLF of model C at various redshifts is
shown in Figure \ref{fig:allrlf}. The broken power-law form for the
whole population can clearly be seen at $z\ltsimeq 1$. At higher
redshifts the continued evolution of the high--luminosity population
causes the RLF to have a small dip at $\log(L_{151}) \approx 26.2$
(26.6 for $\Omega_{\rm M}=0$). This is a consequence of the models
chosen and may or may not be real. No previous low--frequency complete
samples have gone to such low flux-density levels to constrain this region of
the $L_{151}-z$ plane directly. In fact, for $\log(L_{151}) \approx 26.2$ and
$z>1$, the 151 MHz flux-density of any sources would be fainter than the limit
of the 7CRS sample and the RLF is purely constrained by the source
counts in this region. The maximum size of this feature is only 0.2
dex, so it would be extremely difficult to confirm directly by
measurement of the RLF.

\begin{figure}
\epsfxsize=0.47\textwidth
\hspace{-0.4cm}
\epsfbox{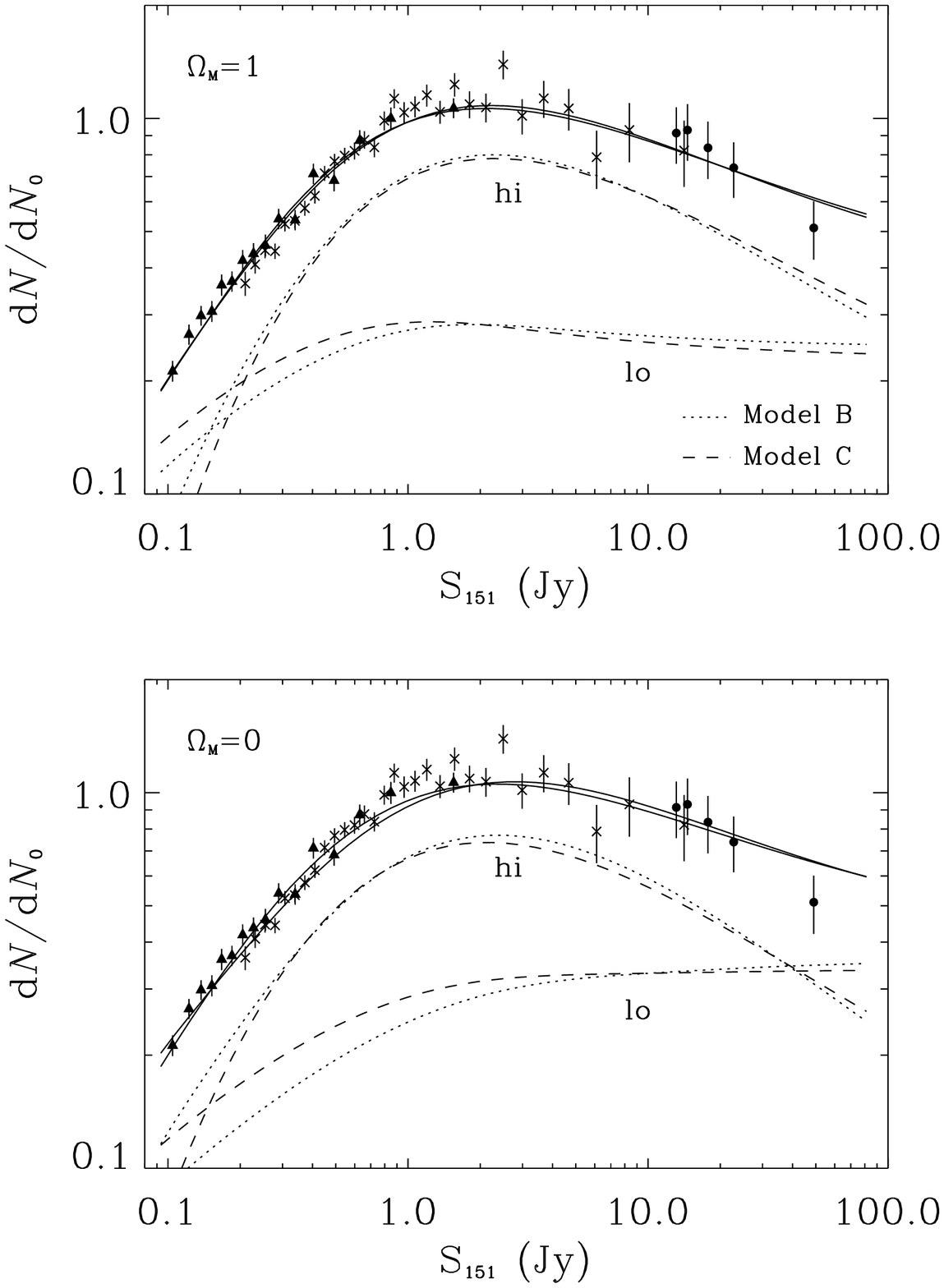}
{\caption[junk]{\label{fig:scects} The 151 MHz source counts data (as
in Fig. \ref{fig:polysc}) with model-B (short-dash) and
model-C (long-dash) fits for $\Omega_{\rm M}=1$
(top) and $\Omega_{\rm M}=0$ (bottom). The contributions
to the source counts from the low--luminosity and
the high--luminosity populations are shown separately (and marked `hi'
and `lo'). The solid lines show the total source counts from both
models. }}
\end{figure}

Fig. \ref{fig:scects} shows the calculated source counts at 151 MHz
from the model C RLFs for $\Omega_ {\rm M}=1$ and $\Omega_ {\rm
M}=0$. The counts from each of the two parts of the RLF have also been
plotted separately. The models fit the total source counts very well,
as is expected because the maximum likelihood method minimises the
$\chi^{2}$ of the fit to the source counts data. For both cosmologies,
the high--luminosity population dominates the source counts over most
of the flux-density range. The low--luminosity population starts to
dominate at $\sim 0.2$ Jy. This suggests that fainter samples, such as
those selected at flux-limits $S_{151}\geq 0.1$ Jy, would
contain more low--luminosity population sources. This idea will be
discussed further in Section \ref{rlfconc}.

\begin{figure}
\hspace{-0.3cm} \epsfxsize=0.47\textwidth \epsfbox{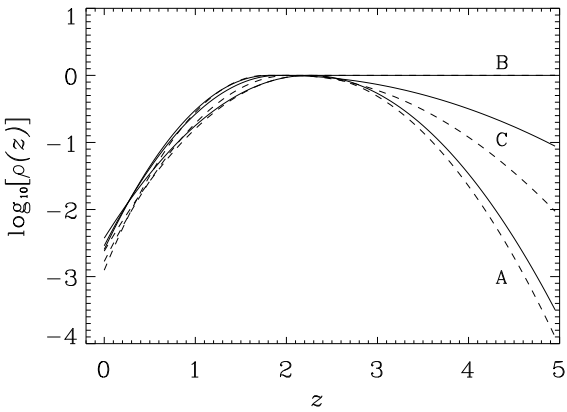}
{\caption[junk]{\label{fig:zdistrlf} The best-fit redshift dependence
of the high--luminosity part of the RLF for the 3 models considered
here ($\Omega_ {\rm M}=1$ -- solid lines, $\Omega_ {\rm M}=0$ --
dashed lines). Note that the curves for model B in the two cosmologies
are the same at high-redshift, so the dashed curve is obscured by the
solid curve.  All are normalised such that $\log_{10} (\rho)=0$ at the
peak redshift.}}
\end{figure}

\begin{table}
\footnotesize
\begin{center}
\begin{tabular}{cccccc}
\hline\hline 

\mc{1}{c}{Model}&\mc{1}{c}{$\Omega_{\rm M}$}&\mc{1}{c}{$N_{\rm par}$} &\mc{1}{c}{$P_{\rm KS}$}\\
\hline\hline

A & $1$ & $10$ & $0.11$  \\ 
B & $1$ & $10$ & $0.28$  \\ 
C & $1$ & $11$ & $0.36$  \\ 
\hline			
A & $0$ & $10$ & $0.03$  \\ 
B & $0$ & $10$ & $0.09$  \\ 
C & $0$ & $11$ & $0.07$  \\
\hline\hline         
\end{tabular}
\end{center}              
{\caption[Table of observations]{\label{tab:rlfgof} Goodness-of-fit
table.  $N_{\rm par}$ is the number of free parameters in the model
(including those from both the low-- and high--luminosity populations) and
$P_{\rm KS}$ is the probability from the 2D K-S test to the
$L_{151}-z$ distribution.  As in Table \ref{tab:rlffit} the top
three rows refer to $\Omega_{\rm M}=1$ and the bottom three to
$\Omega_{\rm M}=0$. }} \normalsize
\end{table}

We now briefly consider whether it is obvious that any of these three
models provides the best-fit to the data and hence whether we can
usefully constrain the strength of any high--$z$ decline in the
co-moving space density $\rho$ of radio sources. We defer a full
statistical analysis of this question to a future paper (Jarvis et
al., 2000) in which data from the 3CRR and 6CE complete samples is
combined with the results of targeted searches for $z > 4$ radio
galaxies (the 6C* sample of Blundell et al. 1998).  Figure
\ref{fig:zdistrlf} shows the redshift-dependence of the
high--luminosity RLF for these three models with their free parameters
fixed at the best-fit values. All the models provide good fits to both
the source counts and local RLF data, with reduced $\chi^{2}$ values
of order one. Table \ref{tab:rlfgof} shows the goodness-of-fit (KS
test probabilities) of the models for the two cosmologies considered.
In both cosmologies the KS test shows that models B and C are
marginally favoured over model A -- the symmetric decline model.  The
extra parameter in model C ($z_{{\rm h}2}$ -- the slope of the decline
in number density at high--redshift) enables the strength of any
high--$z$ decline to be fitted directly.  The fact that the
best-fitting values of $z_{{\rm h}2}$ are greater than those of
$z_{{\rm h}1}$ but less than infinity indicates a moderate high--$z$
decline fits the data best (see Fig.  \ref{fig:zdistrlf}). We
conclude, however, that we cannot rule out with any level of
confidence a constant space density up to $z \sim 5$.

\subsection{Simulated data}
\label{simul}

\begin{figure*}
\epsfxsize=0.95\textwidth
\hspace{0.45cm}
\epsfbox{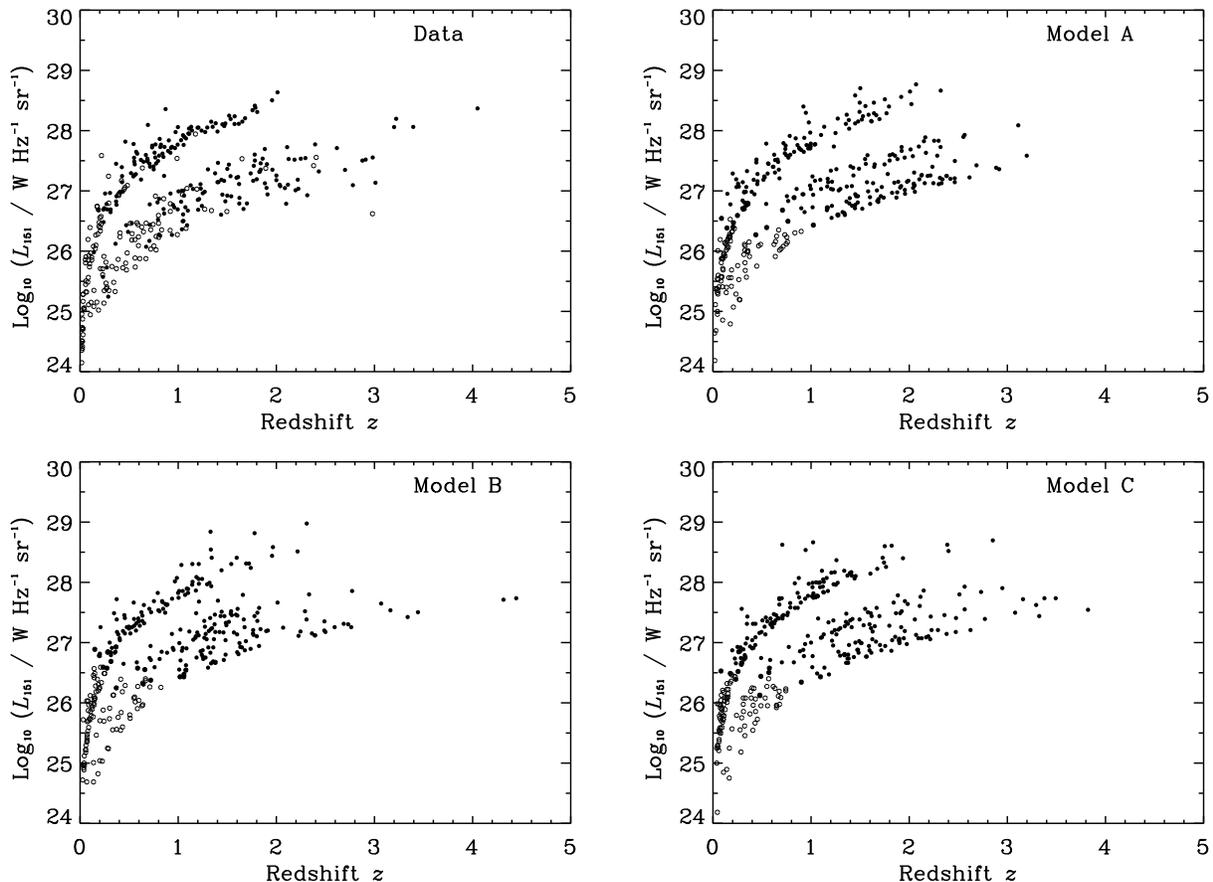}
\vspace{0.2cm} {\caption[junk]{\label{fig:mcrlf} The top-left panel
shows the $L_{151}-z$ plane for the 3CRR, 6CE and 7CRS complete
samples. Objects in the samples with emission line luminosities
$\log_{10} (L_{\rm [OII]} / {\rm W}) \geq 35.1$ are shown as filled
symbols whilst those at lower luminosities are open. The other three
panels show simulations of the $L_{151}-z$ plane for samples with the
same flux-limits and sky areas as the complete samples. These
simulations were generated using the best-fit RLF models A, B and C
($\Omega_{\rm M}=1$). Objects drawn from the low--luminosity population
are shown as open circles and those from the high--luminosity
population are open. }}
\end{figure*}

Simulations of the $L_{151}-z$ plane for the three models are shown in
Figure \ref{fig:mcrlf} along with the actual $L_{151}-z$ data for the
complete samples (for $\Omega_{\rm M}=1$). Statements made in this
section concerning comparison of real and simulated data apply also to
an $\Omega_{\rm M}=0$ cosmology. The first thing to notice is that all
the models reproduce the $L_{151}-z$ data reasonably well. Sources
with emission line luminosities $\log_{10} (L_{\rm [OII]} / {\rm W})
\geq 35.1$ (see Willott et al. 1999, 2000 for discussion of this
adopted value) are shown as filled symbols whereas those with lower
luminosities are open. Note that because the radio galaxies in the
combined sample span a large range of redshift, Balmer lines are often
not observed and a classification scheme such as that adopted by Laing
et al. (1994) to distinguish between low-- and high--ionisation
sources cannot be used. Therefore we make a distinction between the
two populations simply on the basis of emission line luminosity.

For the model simulations, the high and low luminosity populations are
similarly separated into filled and open symbols. The transition
between populations at $\log_{10} (L_{151} /$ W Hz$^{-1}$ sr$^{-1})
\approx 26.5$ is well-reproduced by the models indicating that the two
populations can approximately be separated according to the strength
of their narrow emission lines. As seen in Willott et al. (1999), this
quantity is likely to be closely related to the accretion rate of the
source.  The scatter in the emission line--radio correlation is not
taken account of in the RLF models, which explains why the filled and
open symbols overlap in the real data, but do not for the
simulations. This scatter is due to effects such as the range of radio
source ages and environments which give different radio luminosities
for a given jet power (see Willott et al. 1999; Blundell et al. 1999).

At high--redshifts ($z \approx 3$) the very few numbers of objects in
the samples means that small number statistics become important. For a
comparison with simulated data, the situation is even worse because
both the data and the simulations have independent Poisson errors.
Therefore, the simulations shown in this section should not be used to
distinguish between best-fitting models (a direct integration of the
RLF over redshift is preferable, see Sec.~\ref{sec:zdist}), but these
simulations could give an indication of a poor fit. In any case, the
simulations look very similar, even at $z \approx 3$, for all the
three models tested.

\subsection{Redshift distributions}
\label{sec:zdist}

\begin{figure}
\vspace{0.5cm}
\epsfxsize=0.42\textwidth
\hspace{0.6cm}
\epsfbox{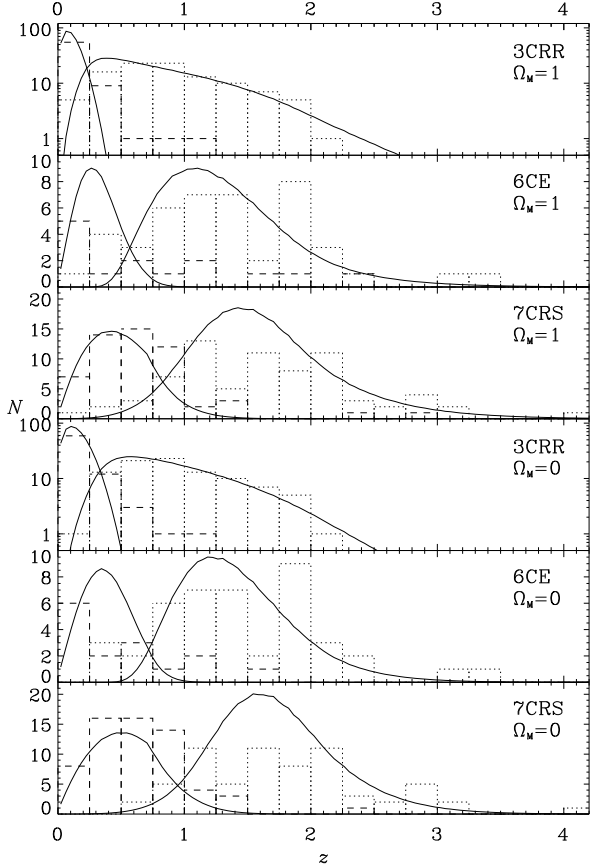}
\vspace{0.5cm} {\caption[junk]{\label{fig:nzmod} The histograms show
the number of sources in the 3CRR, 6CE and 7CRS complete samples
binned in redshift with bin width $\delta z=0.25$ for $\Omega_{\rm
M}=1$ and $\Omega_{\rm M}=0$. There are separate histograms for the
low (dashed) and high (dotted) luminosity populations. The solid lines
show the model C predictions for the number-redshift distribution for
each population. Note that the 3CRR plot only has a log scale on the
y-axis because of the high peak at low--redshift. }}
\end{figure}

We now consider the redshift distributions for the 3CRR, 6CE and 7CRS
complete samples. As these are obtained by integrating the RLF, the
only source of small number statistics come from the data they are
compared with. Figure \ref{fig:nzmod} shows histograms of the number
of sources per bin of width $\delta z=0.25$ for the 3CRR, 6CE and 7CRS
samples separated into two populations depending upon emission line
luminosities as before. For $\Omega_{\rm M}=0$, the high/low
luminosity divide is at $\log_{10} (L_{\rm [OII]} / {\rm W}) = 35.4$.
Also plotted are the model C RLF predictions for each population for
the three samples.

\begin{figure}
\vspace{0.5cm} 
\epsfxsize=0.42\textwidth
\hspace{0.7cm}
\epsfbox{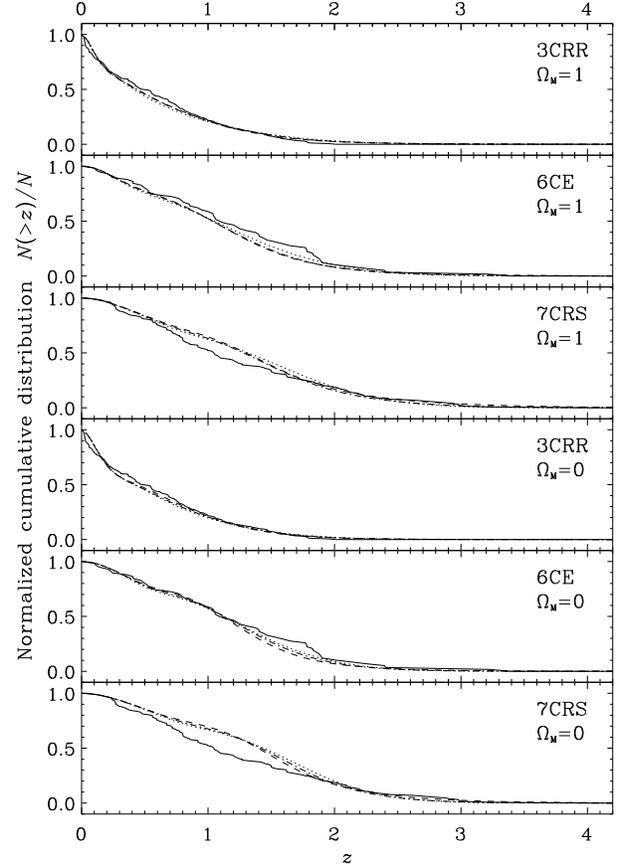}
\vspace{0.5cm} 
{\caption[junk]{\label{fig:nzcumall} Normalised cumulative redshift
distributions of sources in the 3CRR, 6CE and 7CRS complete samples. The
solid curves are the cumulative redshift distributions in the samples
predicted by the RLF of models A (dotted), B (dashed) and C
(dot-dashed).}}
\end{figure}

\begin{figure}
\vspace{0.5cm} 
\epsfxsize=0.42\textwidth
\hspace{0.7cm}
\epsfbox{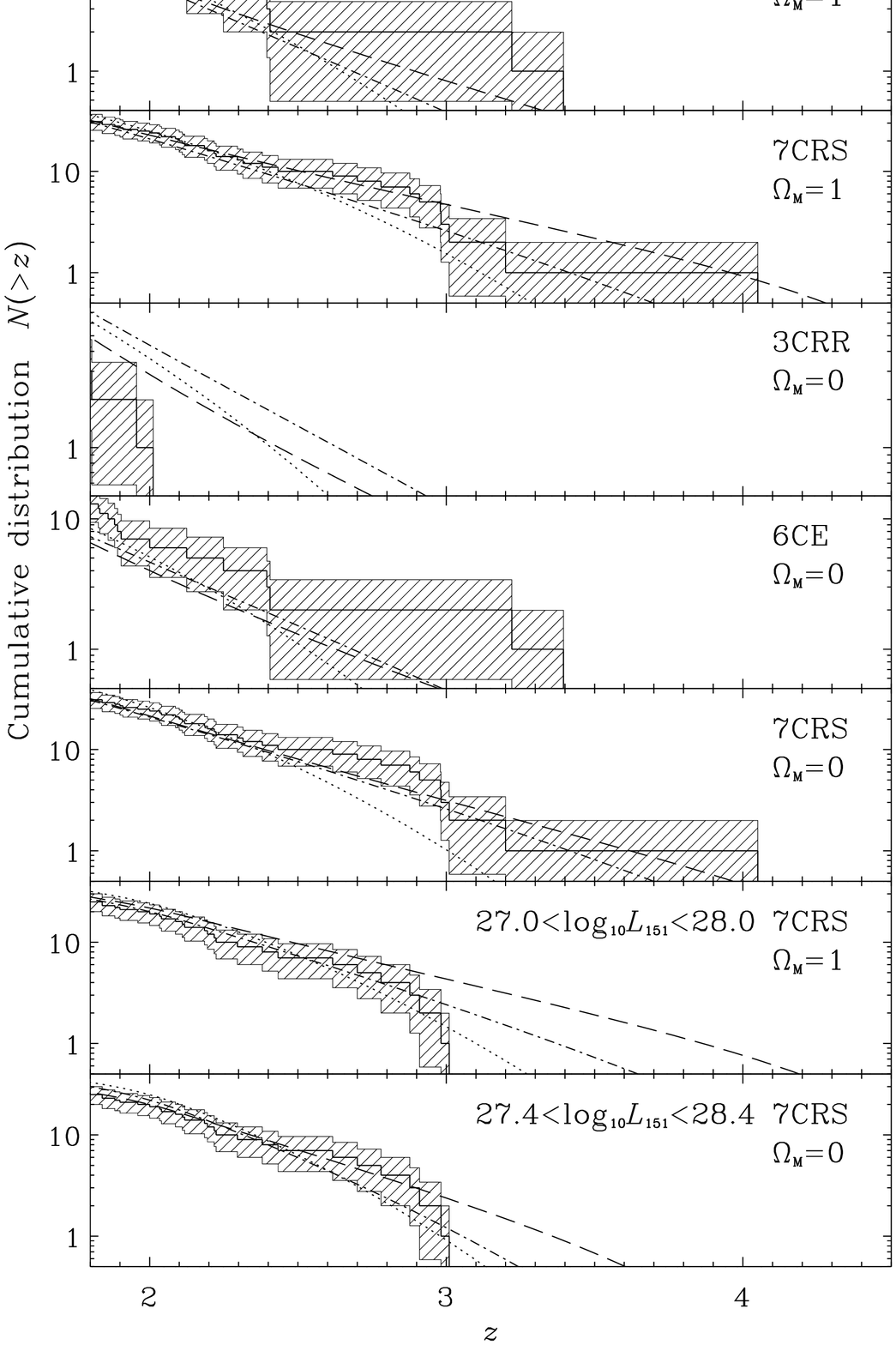}
\vspace{0.5cm} {\caption[junk]{\label{fig:nzdun} Cumulative redshift
distributions at $z>1.8$. Each panel shows the distributions predicted
by each of models A, B \& C as in Fig. \ref{fig:nzcumall}. The data
for each sample is shown as a solid line with the filled region the
$\pm 1 \sigma$ Poisson uncertainty. The top three panels refer to the
3CRR, 6CE and 7CRS samples, respectively, for $\Omega_{\rm M}=1$. The
next three panels show the corresponding plots for $\Omega_{\rm
M}=0$. The final two panels show the 7CRS cumulative redshift
distributions including only sources with luminosities in the range
$27.0 <\log_{10} L_{151} < 28.0$ ($\Omega_{\rm M}=1$) and $27.4
<\log_{10} L_{151} < 28.4$ ($\Omega_{\rm M}=0$).}}
\end{figure}

For the 3CRR sample, the model fits very well, the main exception
being the low luminosity objects at $z >0.3$, where there are several
more sources observed than predicted. This was also seen in the
simulations and attributed to scatter in the emission line--radio
correlation. With only 58 sources in the 6CE sample, the effects of
small number statistics are clear with an irregular observed
distribution. However, in general the model redshift distribution is a
fairly good approximation to the data. There is something of an excess
of low--luminosity sources predicted at low--redshifts ($z<0.5$), but
this can be accounted for by the scatter, since there are several
high--luminosity objects observed at these redshifts.  For the 7CRS
sample, again we find a fairly good fit. The main differences here are
a small deficit predicted at $z\approx 0.5$ and an excess at $z\approx
1.5$. The redshift distributions at high--redshift ($z>2$) will be
discussed later in this section.

Figure \ref{fig:nzcumall} plots the cumulative redshift distributions
of the two samples for all three models. The similarity of these three
curves shows that the models all predict similar redshift
distributions of sources. The one-dimensional KS test was used to test
the difference between the observed and model redshift
distributions. For the 3CRR sample the significances are 0.011 (0.03),
0.007 (0.03) and 0.007 (0.02) for models A, B and C, respectively
($\Omega_{\rm M}=0$ values in brackets). The values for the 6CE sample
are 0.40 (0.32), 0.15 (0.10), 0.16 (0.18) and for the 7CRS sample 0.09
(0.002), 0.04 (0.0004) and 0.09 (0.001). The lowest values of KS
significance are for the 7CRS redshift distribution for the case of
$\Omega_{\rm M}=0$. It is fairly easy to speculate on the cause of
this low probability: the model RLFs appear to be too high at $z \sim
1.5$ and $\log_{10} L_{151} \sim 26.5$ and too low at $z \sim 0.5$ and
$\log_{10} L_{151} \sim 26$ ($\Omega_{\rm M}=1$, for $\Omega_{\rm
M}=0$ these luminosities are $\log_{10} L_{151} \sim 26.9$ and
$26.4$). This is almost certainly a limitation of our modelling
procedure since these luminosities lie close to the break luminosities
$L_{{\rm l} \star}$ and $L_{{\rm h} \star}$ which bracket the range
over which the two populations overlap. Away from these break
luminosities the rates of exponential decline in luminosity of each
population are fixed parameters in our model, and so their combination
lacks the flexibility to fit perfectly all data within the overlap
region. This situation could be improved by using additional free
parameters, for example parameters controlling the rates of
exponential decline, or by use of free-form fits (c.f. DP90).

To compare the numbers of high--redshift objects in the samples with
the models, Fig. \ref{fig:nzdun} plots the un-normalized cumulative
redshift distributions at $z>1.8$ on a logarithmic scale. For the 3CRR
sample, model B (that with no decline in $\rho$ at high--redshifts)
predicts four sources at $z>2$ ($\Omega_{\rm M}=1$; for $\Omega_{\rm
M}=0$ the model predicts three sources) whereas there is only one
observed. Using Poisson statistics the probability of observing one or
less objects given a mean of four is 0.1. Therefore the observations
are marginally incompatible with the model. However, we note that the
lack of $z>2$ objects in the 3CRR sample could also be due to a
steepening of the RLF at $\log_{10} L_{151} \sim 28.5$, due to a
maximum radio luminosity of AGN, perhaps related to the maximum black
hole mass in galaxies at these redshifts. This could be included in
RLF models, but we do not do this in this paper, since there is little
constraining data at such high luminosities and we wish to keep the
numbers of free parameters in our models to a minimum. Note that the
new equatorial sample of powerful radio sources defined by Best,
R\"ottgering \& Lehnert (1999) which has selection criteria producing
a sample very similar to the 3CRR sample does indeed contain four
sources at $z>2$, so the lack of such sources in the 3CRR sample may
well be due to small number statistics. The evolution of the RLF for
such high--luminosity sources is discussed in more detail in Jarvis et
al. (2000).

Moving on to consider the 6CE and 7CRS redshift distributions
(Fig. \ref{fig:nzdun}), we find that all three models are consistent
with the high--$z$ data to within $1 \sigma$ uncertainties. There is
no evidence here for a decline in the co-moving space density of radio
sources at high--redshift.

\begin{figure*}
\vspace{-0.4cm} 
\epsfxsize=0.85\textwidth
\hspace{0.6cm}
\epsfbox{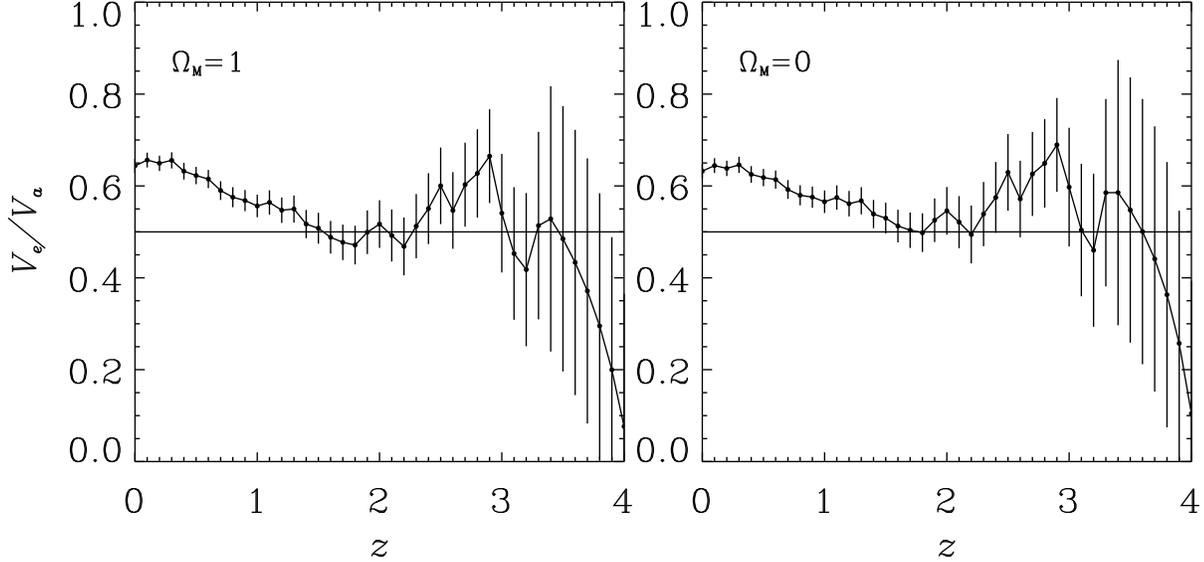}
\vspace{0.3cm} 
{\caption[junk]{\label{fig:vmax} Banded $V/V_{\rm
max}$ test for the 356 sources in the 3CRR/6CE/7CRS complete
samples. The left panel is for $\Omega_{\rm M}=1$ and the right panel
for $\Omega_{\rm M}=0$. The error bars are equivalent to
1$\sigma$. There are no significant differences between the form of
the $V/V_{\rm max}$ tests for the two cosmologies.}}
\end{figure*}

The main benefit of the addition of the 7CRS complete sample to the
existing 3CRR and 6CE samples, is that it enables the evolution in the
luminosity band $27.0 <\log_{10} L_{151} < 28.0$ ($\Omega_{\rm M}=1$)
to be determined over the wide redshift range $z \approx 0.3$ to $z
\approx 3$ (see Fig. \ref{fig:allpz}). DP90 found stronger evidence
for a redshift cut--off in the steep-spectrum population by
considering only the sources within a similar luminosity band
(accounting for the difference in selection frequency). Therefore we
now repeat our analysis using only the sources within this luminosity
band (note that for $\Omega_{\rm M}=0$, the band chosen is that
corresponding to equivalent source flux at $z=2$, i.e.  $27.4
<\log_{10} L_{151} < 28.4$). The bottom two panels of
Fig. \ref{fig:nzdun} repeat the 7CRS cumulative redshift
distributions, but now including only those sources in these
luminosity bands. For $\Omega_{\rm M}=1$, the no cut--off model B now
appears inconsistent with the $z>3$ data. The model predicts four
sources at $z>3$ whereas only one is observed. This gives the same
probability as we found with the $z>2$ 3CRR sources, i.e. a
probability of 0.1 that the model and data are consistent. For the
case of $\Omega_{\rm M}=0$, only two sources are predicted at $z>3$ by
the model, so this is certainly consistent with the observation of one
source.

Thus we find very marginal evidence (90\% confidence level) that the
numbers of objects in the $27.0 <\log_{10} L_{151} < 28.0$ luminosity
band falls slightly below the no cut--off model prediction. There is
however, one systematic effect which might reduce even this
significance.  The expected numbers of high--redshift sources were
calculated from the models assuming a spectral index between 151 MHz
and $151 \times (1+z)$ MHz of $\alpha=0.8$.  However, if high-$z$
sources were systematically significantly steeper than $\alpha=0.8$,
then the expected number of high-$z$ sources would be smaller. This is
because a source of a given luminosity on the flux-limit with $\alpha
= 0.8$, would clearly fall below the flux-limit if $\alpha$ were
greater than this (see figure 1 in Blundell et al. 1999).

Our studies of complete samples (Blundell et al. 1999) do not reveal
any intrinsic correlation between redshift and $\alpha$ evaluated at
151\,MHz ($\alpha_{151}$), however the correlation between luminosity
and $\alpha_{151}$ will mean that any high-redshift sources (which
must be high luminosity) will tend to have steeper spectral indices.
The consequences of this effect are discussed in detail in
Section 3 of Jarvis \& Rawlings (2000) and we defer interested readers
to that paper. Re-calculation assuming $\alpha=1.0$ at high--redshifts
for model B gives an expectation of only two sources at $z>3$, thereby
removing any discrepancy between this model and the data. Finally, we
note that the chosen luminosity band of $27.0 <\log_{10} L_{151} <
28.0$ excludes the four highest redshift objects in the combined
sample (see Fig. \ref{fig:allpz}) which have luminosities just above
this limit. Extending the luminosity range considered to $27.0
<\log_{10} L_{151} < 28.5$ would give a similar result to those found
for the entire samples.

\section{The $V/V_{\rm max}$ test}
\label{vvmax}

In this section the $V/V_{\rm max}$ test is applied to the complete
sample data. This test calculates the ratio of the volume $V$ in the
Universe enclosed by each source of redshift $z$ to the maximum volume
$V_{\rm max}$, corresponding to the maximum redshift $z_{\rm max}$ a
source of this luminosity would have if its flux-density lies above
one or more of the survey flux-limits. If there is no cosmic evolution
then sources would have values of $V/V_{\rm max}$ randomly distributed
between 0 and 1 (Rowan-Robinson 1968; Schmidt 1968).  Hence by
calculating the values of $V/V_{\rm max}$ for all the sources in a
sample and averaging, one can determine the mean evolution of the
sample. A value of $<V/V_{\rm max}>$ greater than 0.5 corresponds to
positive redshift evolution and less than 0.5 indicates negative
evolution. The distribution of $<V/V_{\rm max}>$ expected in the case
of no evolution is, for a large sample size $N$, a Gaussian centred on
0.5 with $\sigma= (12N)^{-1/2}$.

Avni \& Bahcall (1980) devised a method for combining samples with
various flux-limits and sky areas for the application of a generalised
$V/V_{\rm max}$ test. This version combines several samples such that
they can be treated as a single sample with a variable sky area
depending upon the flux-limit. The new test variables are the volume
enclosed by sources $V_{\rm e}$ and the volume available in the new
sample $V_{\rm a}$. Note also that the fact that the 6CE complete
sample has an upper as well as a lower flux-limit means that the
volume available for a source in the 6CE sample is given by $V_{\rm a}
=V_{\rm max}- V_{\rm min}$. To calculate values of $V_{\rm a}$ one
must make some assumption about the shapes of the radio spectra to
determine values of $z_{\rm max}$. For this test it is assumed that
radio spectra are not significantly curved over the region of
interest, and spectral indices at a rest-frame frequency of 151 MHz
are used.

The known strong positive evolution from $z=0$ to $z\approx 2$ would
completely mask any change in the evolution at higher redshifts in the
standard $V/V_{\rm max}$ test. Therefore a banded version of the test
is used here (e.g. Osmer \& Smith 1980). This modification calculates the
mean $V_{\rm e}/ V_{\rm a}$ restricting the analysis to $z>z_{0}$ at
several different values of $z_{0}$. Hence all the evolution below
$z_{0}$ is masked out of the analysis. The statistic is now
\begin{equation}
\frac{V_{\rm e}-V_{0}}{V_{\rm a}-V_{0}},
\end{equation}
where $V_{0}$ is the volume enclosed at redshift $z_{0}$. 

\begin{figure}
\epsfxsize=0.43\textwidth
\hspace{0.59cm}
\epsfbox{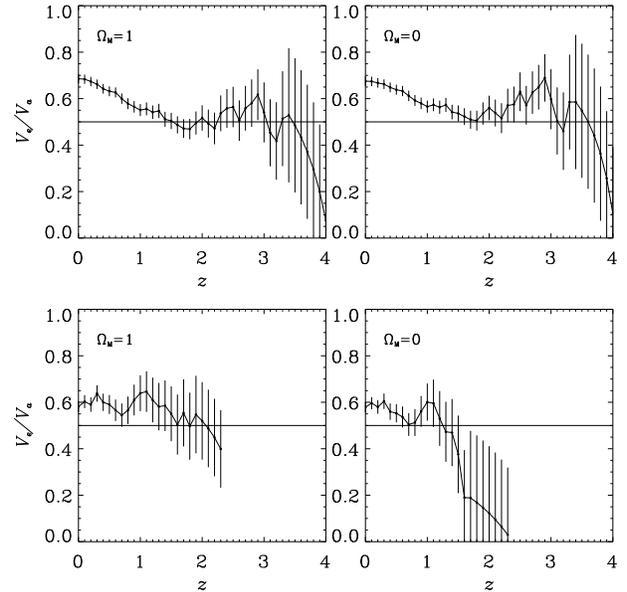}
\vspace{0.3cm} {\caption[junk]{\label{fig:vmaxpop} Banded $V/V_{\rm
max}$ test for the high-- (top panels) and low-- (bottom panels)
luminosity populations separately. The left panels are $\Omega_{\rm
M}=1$ and the right panels $\Omega_{\rm M}=0$.}}
\end{figure}

Fig.~\ref{fig:vmax} shows this banded $V/V_{\rm max}$ test for the
complete samples. In the banded test, the adjacent points are not
statistically independent and the error bars are much larger than the
typical dispersion of points. Note that at low--redshifts ($z<1.5$)
the strong positive evolution of the radio source population is
clearly seen with the values of $V_{\rm e}/V_{\rm a}$ significantly
above the 0.5 no-evolution line. At $z=1.5$ the values approach the
0.5 line where they remain (within the errors) out to the redshift
limit of this sample ($z=4.0$). The positive bump at $z\approx 2.8$ is
marginally significant, but at this point there are very few sources
left in the sample, so no firm conclusions can be drawn from
this. Thus there is no evidence from the $V/V_{\rm max}$ test for a
high--$z$ decline (or increase) in the comoving space density of radio sources.
\begin{figure}
\epsfxsize=0.43\textwidth
\hspace{0.59cm}
\epsfbox{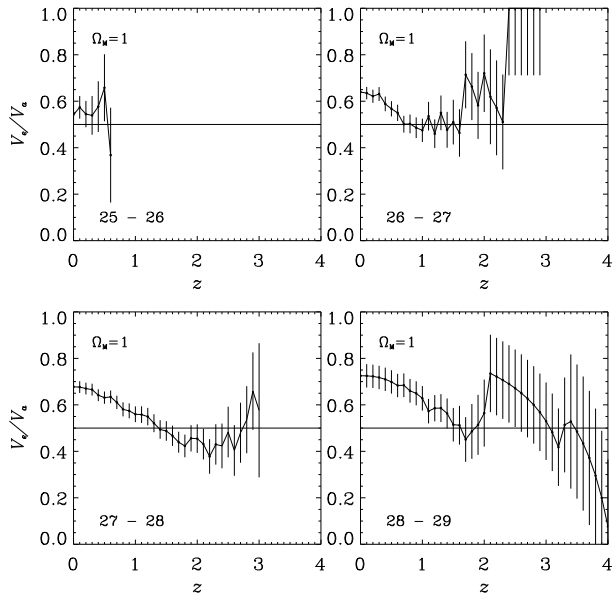}
\vspace{0.3cm} {\caption[junk]{\label{fig:vmaxlum} Banded $V/V_{\rm
max}$ test for the complete samples separated into sources in
different luminosity ranges. The ranges in $\log_{10} L_{151}$ included
in each panel are labelled in the bottom-left corners ($\Omega_{\rm
M}=1$ only is shown here).}}
\end{figure}

The $V/V_{\rm max}$ test can also be used to study the evolution of
the low-- and high--luminosity populations separately. The complete
samples were separated into two groups, depending upon their emission
line luminosities as described in Sec. \ref{simul}. These groups are supposed
to signify the division between objects in the low-- and
high--luminosity populations in the RLF. The banded $V/V_{\rm max}$
test was performed for each group and the results shown in
Fig. \ref{fig:vmaxpop}. Both populations undergo positive evolution at
low--redshift ($z<0.5$) with the evolution stronger (and at a higher
statistical significance) for the high--luminosity sources. At $z>1$
most sources in the complete samples belong to the high--luminosity
population, so its $V/V_{\rm max}$ values are essentially the same as
those in Fig. \ref{fig:vmax} for the whole dataset. There are hints of
continuing positive evolution in the regime $0.5<z<1$ for the
low--luminosity sources (this is more apparent for $\Omega_{\rm M}=1$
than $\Omega_{\rm M}=0$). This evolution of the low--luminosity
population is consistent with that derived in the model RLFs, where
the positive evolution ceases at $z\approx 0.7$.

In Fig. \ref{fig:vmaxlum} we show the results of separating the sample
into luminosity bands and repeating the $V/V_{\rm max}$ test. The
$0<z<1$ evolution appears to depend strongly on luminosity. The
luminosity separation does not provide any evidence for negative
evolution at high--redshifts (i.e. a redshift cut--off) in either
of the two high--luminosity bands. Therefore we conclude from this
Section that the $V/V_{\rm max}$ test shows positive evolution at
$z<2$ which depends strongly on luminosity, but is inconclusive at
higher redshifts.

\section{Comparison with previous work} 
\label{comprlf}

\subsection{Comparison with DP90}

\begin{figure}
\epsfxsize=0.48\textwidth
\epsfbox{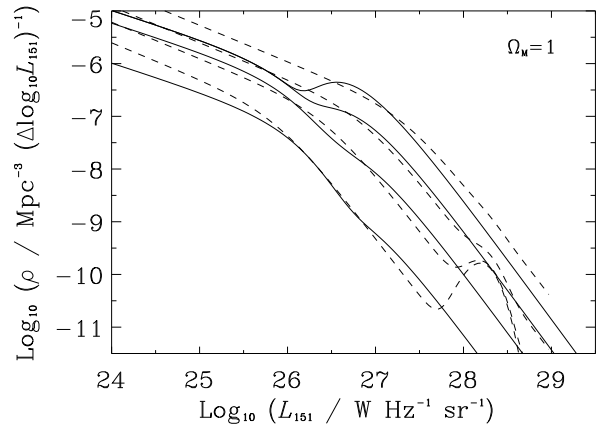}
{\caption[junk]{\label{fig:dprlf} The evolving radio luminosity
function of our model C (solid lines) plotted along with the PLE model of
DP90 (dashed lines). The curves correspond to $z=0, 0.5, 1, 2$
(bottom to top). The DP90 model has been transformed to 151 MHz by
assuming all sources have spectral indices of 0.8.}}
\end{figure}

In Fig. \ref{fig:dprlf}, our model-C RLF is plotted with the
steep-spectrum PLE model of DP90 at various redshifts ($\Omega_{\rm
M}=1$ only). The DP90 LDE model which incorporates negative luminosity
evolution at high--redshift is not shown here, since it is almost
identical to the PLE model in the redshift range $0.5<z<3$ (which is
to be expected as these regimes are constrained by data). To account
for the fact that DP90 evaluated the RLF at high--frequency (2.7 GHz),
their luminosity scale has been transformed to 151 MHz assuming a
global spectral index of 0.8. The kinks in the DP90 model at
$\log_{10} (L_{151} /$ W Hz$^{-1}$ sr$^{-1}) \approx 28.2$ are due to
the use of a seven term power-law expansion for the non-evolving
low--luminosity RLF component: there are no such powerful sources at
low--redshift in their samples (or indeed in the 3CRR sample) and
hence no constraint from the data, so this feature is a numerical
artefact. As described in Sec.~\ref{sec:zdist} our modelling procedure
is also imperfect: the bumps in our estimations of the RLF around the
breaks in the high-- and low--luminosity populations may be artefacts
of the restricted number of free parameters in our models.

In general, the DP90 PLE model and model C agree reasonably well. At
low luminosities the PLE model has a steeper slope of 0.69 than the
slope of model C (0.54). Again, this slope is only well constrained at
low--redshifts and the cause of a differing slope is probably due to
the use here of a more recent determination of the local radio
luminosity function. The break luminosity appears at exactly the same
value for the two models. Note how our use of two populations with
differential evolution mimics the evolution of the break luminosity in
the PLE case. Since we know that radio sources are short-lived with
respect to the Hubble Time, pure luminosity evolution has no physical
meaning and there is clearly not one population of radio sources which
fades over cosmic time. In contrast, the two population model has some
physical meaning and mimics PLE by having only {\em density} evolution
for both populations.

The high--luminosity power-law slopes for the DP90 PLE model and our
model C are very similar. However, at $z=2$ there is a clear excess of
high--luminosity ($\log_{10} L_{151} > 27.5$) sources in the DP90
model. At $\log_{10} L_{151}=28$ the DP90 RLF is a factor of 2.1
higher than that of our model C. At $z=3$ and $\log_{10} L_{151}=28$
this factor has increased to 2.4. The effect of this is shown by a
comparison of the expected high--redshift sources in the 6CE sample
for RLF models with no redshift cut--off in Dunlop (1998) and in this
paper (Fig. \ref{fig:nzdun}). The plot by Dunlop shows a predicted
excess at $z>2$ over the number observed and this excess continues
beyond the last data point at $z\approx 3.4$. Dunlop takes this as
further evidence for the redshift cut--off. However, in
Fig. \ref{fig:nzdun} (second panel from top) we see that the 6CE
redshift distribution is consistent with our model B at all
redshifts. How can these two models, both of which claim to be no
cut--off models, have such different predictions for the expected
numbers of high--redshift sources? The answer is that all no cut--off
models simply take the RLF at its peak at $z\approx 2$ and `freeze'
the shape and normalization up to high--redshifts (at least to $z \sim
5$). Therefore the DP90 models naturally predict more of these
luminous sources at high redshifts because they have more of them at
$z\approx2$.

The next question is why do DP90 find a higher density of powerful
sources at $2 \ltsimeq z \ltsimeq 3$ than found by our study? The
answer is probably due to the fact that a very high fraction of the
most distant sources in DP90 did not have reliable redshifts:
inspection of fig. 9 of DP90 shows that only a quarter of their $z>2$
steep-spectrum PSR sources had spectroscopic redshifts with the
remainder estimated from the $K-z$ diagram. Dunlop (1998) reports that
further spectroscopy has shown the redshifts in DP90 to be typically
over-estimated (most likely due to the positive correlation between
radio luminosity and $K$-band luminosity observed by Eales et
al. 1997). If these sources are actually at systematically lower
redshifts then this explains why DP90 observe an excess of $z>2$
sources. Extrapolating this high source density up to $z=5$ then
predicts many sources in no cut--off models. Note that because few of
their redshift estimates were actually in the region of supposed
decline (at $z>3$) the redshift over-estimation has the opposite
effect to the naive expectation that this new information strengthens
claims for the redshift cut--off. Thus we have an explanation as to
why DP90 find stronger evidence for a steep-spectrum redshift cut--off
than we find with the 3CRR, 6CE and 7CRS datasets (having
virtually complete spectroscopic redshifts).

DP90 argued that their $V/V_{\rm max}$ test on the steep-spectrum
population showed clear evidence for a redshift cut--off beyond
$z=2$. However we do not find a similar behaviour in our low--frequency
selected samples, with apparently no evolution between $z=2$ and
$z=4$. Inspection of figs. 12 and 13 of DP90 show that their evidence
of negative density evolution at $z \gtsimeq 2$ is highly tentative:
their data are only about 1$\sigma$ away from the null-hypothesis of
no evolution. The banded $V/V_{\rm max}$ test suffers significantly
from large statistical errors when only a few sources are left in the
samples at the highest redshifts, and this tends to limit its
usefulness in practice. Combined with the uncertainty in the DP90
redshift estimates, it is difficult to see clear evidence for a
redshift cut--off in their $V/V_{\rm max}$ tests.

\subsection{Comparison with the radio-loud quasar RLF}

\begin{figure}
\epsfxsize=0.48\textwidth
\epsfbox{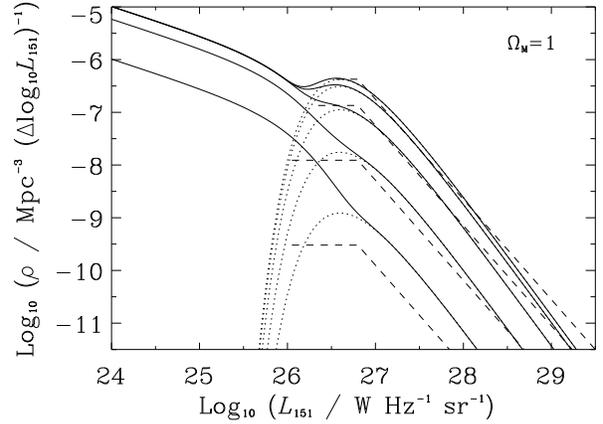}
{\caption[junk]{\label{fig:qrlfall} The evolving radio luminosity
function of model C (solid lines) plotted along with the quasar RLF
model C from W98 (dashed lines). The quasar RLF has been multiplied by
a factor of 2.5 to account for the fraction of luminous quasars
compared to radio galaxies in low--frequency selected samples
above the critical luminosity $\log_{10} L_{151} \approx 26.5$
(Willott et al. 2000). Also plotted as dotted lines
is the high--luminosity part of the model-C RLF. The curves
correspond to $z=0, 0.5, 1, 3, 2$ (bottom to top).}}
\end{figure}

In line with unified schemes (e.g. Antonucci 1993), the
high--luminosity part of the RLF derived here consists of radio
galaxies and quasars which are identical except for the angle our
line-of-sight makes with the jet axis. In Willott et al. (2000) we
showed that the fraction of quasars in the high luminosity population
is 0.4.  Therefore one would expect a very similar RLF to that of the
quasars in W98, with an increase in number density of $\approx
2.5$. Fig. \ref{fig:qrlfall} plots the RLF derived for all radio
sources in this paper and the quasar RLF from W98 multiplied by a
factor of 2.5. At high--redshifts the two RLFs appear very similar
(there are very few low--redshift/luminosity quasars in the complete
samples -- see Willott et al. 2000 for a discussion of this). The
high--luminosity power-law slope for model C is 2.3, compared with 1.9
for the quasar RLF. The peak for the high--luminosity Schechter
function is at approximately the same luminosity ($\log_{10} (L_{151})
\approx 26.6$) as the break required by the source counts for the
quasar RLF. The quasar RLF of W98 had a flat slope below this break,
whereas the models of this paper have very rapid declines with
decreasing luminosity.  If the models described here are correct, then
the quasar RLF should also have this Schechter function form. A survey
at fainter flux-limits would probe the break region at high--redshift
and hopefully constrain the number of quasars (or BLRGs) at lower
radio luminosities. A sample of quasars with a flux-limit of
$S_{151}\geq 0.1$ Jy has already been defined by Riley et al. (1999),
however incompleteness arising from the bright optical limit of this
sample (as discussed by W98) will complicate attempts to make
quantitative estimates of the shape of the RLF at and below the break
luminosity.

\section{Discussion}
\label{rlfconc}

Using low--frequency selected radio samples with virtually complete
redshift information, we have investigated the form of the radio
luminosity function (RLF) and its evolution. Our results are generally
in good agreement with previous work by DP90 (Fig.\ref{fig:dprlf}) and
W98 (Fig. \ref{fig:qrlfall}), and most discrepancies can be explained
by limitations of the modelling methods forced on this and previous
studies by the sparse sampling of the $L-z$ plane afforded by
available redshift surveys of radio sources. For the reasons outlined
in Sec.~\ref{sec:intro}, most notably our virtually complete
spectroscopic redshifts, we believe our estimates of the
steep-spectrum RLF to be the most accurate yet obtained.

We find that a simple dual-population model for the RLF fits the data
well, requiring differential density evolution (with $z$) for the two
populations.  This behaviour mimics the effects of a pure luminosity
evolution (PLE) model but with a more plausible physical basis.  The
low--luminosity population can be associated with radio galaxies with
weak emission lines, and includes sources with both FRI and FRII radio
structures.  The comoving space density $\rho$ of the low--luminosity
population rises by about one dex between $z \sim 0$ and $z \sim 1$
but cannot yet be meaningfully constrained at higher redshifts.  The
high--luminosity population can be associated with radio galaxies and
quasars with strong emission lines, and consists almost exclusively of
sources with FRII radio structure.  The comoving space density $\rho$
of this population rises by nearly three dex between $z \sim 0$ and $z
\sim 2$ but cannot yet be meaningfully constrained at higher
redshifts.

Our results on the RLF mirror the situation seen in X-ray and
optically-selected samples of AGN. Low luminosity objects exhibit a
gradual [$\propto (1+z)^{\sim 3.5}$ from $z \sim 0$ to $z \sim 1$] rise
in $\rho$ with $z$ which matches the rise seen in the rate of global
star formation (e.g. Boyle \& Terlevich 1998; Franceschini et
al. 1999), and also the rise in the galaxy merger rate (e.g. Le Fevre
et al. 2000). The $\rho$ of high--luminosity objects rises much more
dramatically [$\propto (1+z)^{\sim 5.5}$ from $z \sim 0$ to $z \sim
2$]. The similarities between the cosmic evolution of radio sources
and these different types of objects have also been pointed out by
Wall (1998) and Dunlop (1998).

We have re-investigated the question of whether there is direct
evidence for a high--redshift decline in the comoving space density
$\rho$ of steep-spectrum radio sources. Applying the $V/V_{\rm max}$
test to the 3C/6CE/7CRS dataset provides no evidence for any such
decline, which is a different result from that obtained by DP90.  A
fundamental limitation is that both our complete samples and those
previously studied by DP90 include only a few sources with $z>2.5$
(e.g. just 12 in 3C/6CE/7CRS), so small number statistics are a huge
problem in assessing the evidence for or against a redshift
cut--off. This is considered in detail in Jarvis et al. (2000).

\begin{figure}
\epsfxsize=0.48\textwidth
\epsfbox{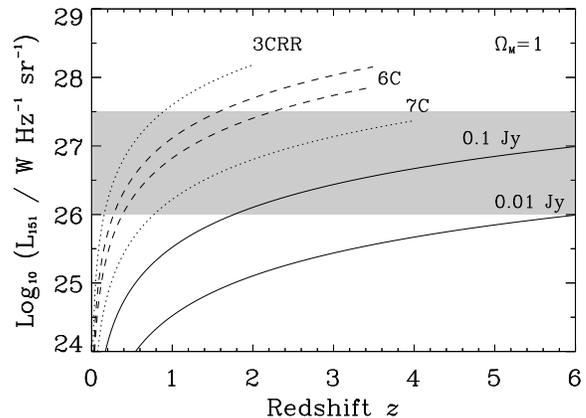}
{\caption[junk]{\label{fig:pzend} The flux-limits on the radio
luminosity -- redshift plane for low--frequency selected samples. The
shaded band shows the luminosity range near the break in the RLF which
contains 70 \% of the luminosity density at high--redshifts.}}
\end{figure}

\begin{figure*}
\vspace{-1cm}
\epsfxsize=0.9\textwidth
\hspace{1.3cm}
\epsfbox{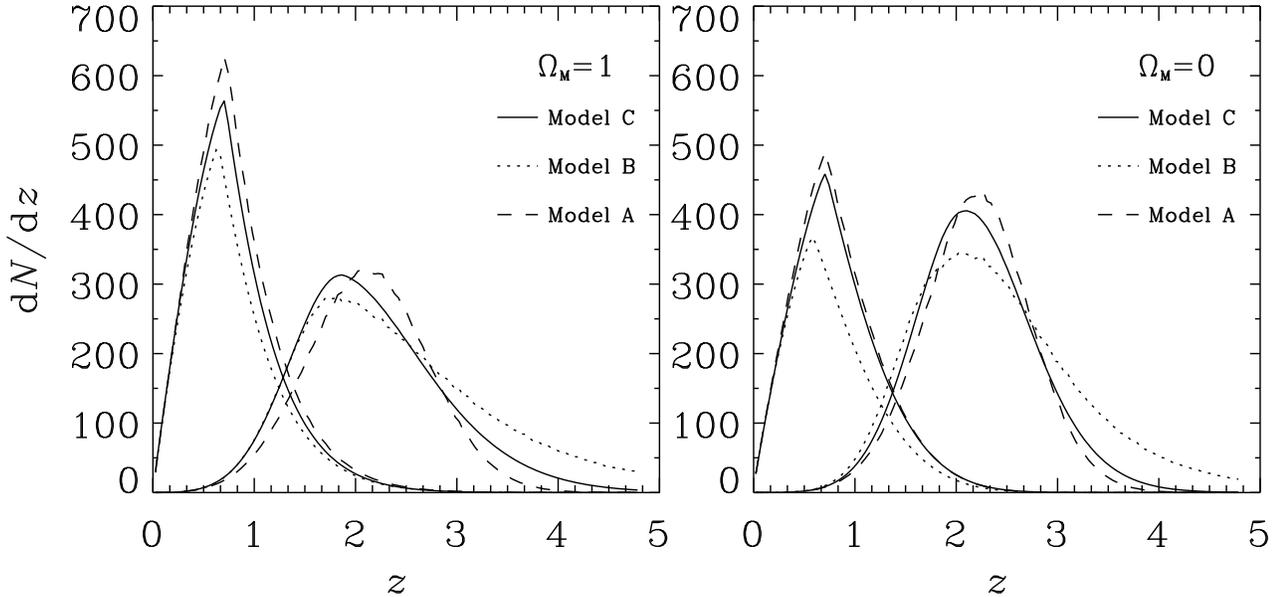}
\vspace{0.3cm} {\caption[junk]{\label{fig:nz0_1} The predicted
redshift distribution of a sample selected at a radio flux-limit of
$S_{151}\geq 0.1$ Jy (five times lower than the 7CRS) for $\Omega_{\rm
M}=1$ (left) and $\Omega_{\rm M}=0$ (right). The low and high
luminosity populations are plotted separately, each for the three
models A, B and C. The vertical axis is the number of sources per unit
redshift. The plots are normalised such that the total number of
sources in the sample equals 1000.}}
\end{figure*}

What would be the best way of clearing up this uncertainty regarding
the redshift cut--off? Larger samples are clearly needed, but in the
case of the bright samples like 3CRR, there is insufficient observable
Universe to improve much on the data already available. Significant
progress on fainter radio samples is however achievable.  The choice
of the flux-density level at which to concentrate efforts for the next
large redshift survey is clearly crucial. We believe that the vital
factor is being able to efficiently probe objects at or just above the
break in the high--luminosity RLF at high--redshift. The reason for
this is that integration of $\rho \times L_{151}$ over the RLFs
plotted in Fig. \ref{fig:allrlf} shows that about $70 \%$ of the
luminosity density of the radio population at $z \sim 2-3$ lies in the
narrow luminosity range $26 \leq \log_{10} L_{151} \leq 27.5$. A
future large redshift survey should be targeted at probing this
population, which is seen only to $z \sim 1$ in the 7CRS.
Fig. \ref{fig:pzend} shows flux-limits in the $L_{151} - z$ plane for
various samples with this luminosity range shaded. The plot
illustrates that $S_{151} \sim 0.1$ Jy is therefore the natural choice
for a new large redshift survey.  Fainter surveys would be less useful
as a probe of the total luminosity density at high--redshift:
inspection of Fig. \ref{fig:pzend} shows that they would become
completely dominated by the low--luminosity population,

To investigate the likely content of a $S_{151} > 0.1$ Jy redshift
survey, we have used models A, B and C RLFs to predict redshift
distributions for a hypothetical 1000-source survey
(Fig.~\ref{fig:nz0_1}).  Note that all models assume that the
evolution of the low--luminosity population freezes at $z \sim 1$.  The
first thing to notice is that the low--luminosity population dominates
the total number of predicted sources in the case of $\Omega_{\rm
M}=1$, whereas in the $\Omega_{\rm M}=0$ case the two populations
become approximately equal. This same result can be seen in the source
counts plots of Fig. \ref{fig:scects}, and arises because the
low--luminosity RLF is currently constrained only at low--redshift, and
the ratio of high-- to low--redshift available volume is much larger in
a low--$\Omega_{M}$ Universe. Thus, although a large fraction of the
sources will be at $z<1$, we predict that in a low--$\Omega_{M}$
Universe in which evolution of the low--luminosity population freezes
at $z \sim 1$, of order 50 \% of the sources will be in the
high--luminosity population and lie at high--redshift, irrespective of
the strength of the redshift cut--off. This should make determination
of the evolution in the luminosity density a rather straightforward
process.  There should be no longer be a problem of small number
statistics, at least at $z \sim 3$.  Adopting model C, 7\% (5\%) of a
hypothetical 1000-source redshift survey would lie at $z\geq3$ for
$\Omega_{\rm M}=1$ (or $\Omega_{\rm M}=0$), whereas the same model
predicts 3\% (2\%) for a survey like the 7CRS.

The pronounced high--redshift tail in the case of `no-cut--off' model B
in Fig.~\ref{fig:nz0_1} should be easily refutable with a 1000-source
redshift survey at $S_{151} > 0.1$ Jy. For the symmetric-decline model
A (and $\Omega_{\rm M}=1$) one would expect about 35 $z\geq3$ sources
in such a sample, whereas for model B we would expect 130.  Indeed,
Dunlop (1998) has already claimed to have ruled out the no-cut--off case
using indirect constraints on the redshift distribution of a LBDS
(Leiden-Berkeley Deep Survey) sample flux-limited at $S_{1.4} \sim 2$
mJy, or $S_{151} \sim 10$ mJy. We note from Fig. \ref{fig:pzend} that
such a sample will be overwhelmingly dominated by the low--luminosity
population, and hence is too deep to provide optimal constraint on the
evolution of the luminosity density of the RLF. Although, as we shall
argue below, indirect evidence favours a gentle decline in $\rho$ at
high--redshift, Dunlop's argument in favour of a redshift cut--off is
not water--tight.  It is currently predicated on an extrapolation
between two samples, the PSR and the LBDS sample, neither of which is
close to spectroscopic completeness, and which are separated by two
orders of magnitude in flux-density; at high--redshift the PSR and LBDS
sources lie on either side of the range in $L_{151}$ which is key to
delineating the evolution of the luminosity density of the radio
source population with redshift.

An alternative approach to obtaining complete redshift information for
large samples is to refine the sample selection criteria to exclude
the large numbers of low--redshift sources. This is being attempted by
several groups using spectral index and/or linear size criteria
(e.g. Bremer et al. 1998; Blundell et al. 1998; Rawlings et al. 1998;
De Breuck et al. 2000). However, the numbers of high--$z$ sources
which are missed in these samples are uncertain, so they will always
struggle to prove the existence of a redshift cut--off, although they
can place a lower limit on the space density (Jarvis et al. 2000).

Another alternative approach is to study the evolution of
flat-spectrum radio sources. Jarvis \& Rawlings (2000) have recently
critically re-evaluated evidence in this area (Peacock 1985; DP90;
Shaver et al. 1996; Shaver et al. 1999) and concluded that
constant-$\rho$ models for the most luminous flat-spectrum sources can
be ruled out at the $\sim 2 \sigma$ level. They favour a decline in
$\rho$ for these objects by a factor $\sim 3$ between $z \sim 2.5$ and
$z \sim 5$, i.e.  behaviour intermediate between models B and C
considered in this paper. However, they also point out that it is not
yet clear how this decline should be interpreted in the context of the
steep-spectrum population. 
\section{Conclusions}

\begin{itemize}
\item We have used three low-frequency selected surveys with essentially
complete spectroscopic redshift information, giving unprecedented
coverage of the radio luminosity versus $z$ plane, to derive the most
accurate measurement of the steep-spectrum RLF yet made.

\item We find that a simple dual-population model for the RLF fits the data
well, requiring differential density evolution (with $z$) for the two
populations.

\item The low--luminosity population can be associated with radio
galaxies with weak emission lines, and includes sources with both FRI
and FRII radio structures; its comoving space density $\rho$ rises by
about one dex between $z \sim 0$ and $z \sim 1$ but cannot yet be
meaningfully constrained at higher redshifts.  The high--luminosity
population can be associated with radio galaxies and quasars with
strong emission lines, and consists almost exclusively of sources with
FRII radio structure; its $\rho$ rises by nearly three dex between $z
\sim 0$ and $z \sim 2$.

\end{itemize}

\section{Acknowledgements}
 
Thanks to Gary Hill, Julia Riley and David Rossitter for their
important contributions to the 7C Redshift Survey.  Thanks also to
Matt Jarvis and Jasper Wall for interesting discussions. We thank the
referee Jim Dunlop for a helpful referees report. We acknowledge the
use of the UKIRT, the WHT and the NRAO VLA which have together made
this project possible. The United Kingdom Infrared Telescope is
operated by the Joint Astronomy Centre on behalf of the U.K. Particle
Physics and Astronomy Research Council. The William Herschel Telescope
is operated on the island of La Palma by the Isaac Newton Group in the
Spanish Observatorio del Roque de los Muchachos of the Instituto de
Astrofisica de Canarias. The Very Large Array is operated by The
National Radio Astronomy Observatory which is a facility of the
National Science Foundation operated under cooperative agreement by
Associated Universities, Inc. This research has made use of the
NASA/IPAC Extra-galactic Database, which is operated by the Jet
Propulsion Laboratory, Caltech, under contract with the National
Aeronautics and Space Administration. CJW thanks PPARC for support.

\end{document}